%% file: WAT_paper.tex
\documentclass[twocolumn,superscriptaddress,showpacs,prd,aps,amsmath,amssymb,nofootinbib]{revtex4}
\usepackage{graphicx}
\usepackage{amssymb}

\input kigou.tex

\begin{document}  

\title{
Non-conformally flat initial data for binary compact objects 
} 

\author{K\=oji Ury\=u}
\affiliation{
Department of Physics, University of the Ryukyus, Senbaru, 
Nishihara, Okinawa 903-0213, Japan}
\author{Fran\c{c}ois Limousin} 
\affiliation{
Laboratoire Univers et Th\'eories, UMR 8102 du CNRS,
Observatoire de Paris, Universit\'e Paris Diderot, F-92190 Meudon, France}
\author{John L. Friedman} 
\affiliation{
Department of Physics, University of Wisconsin-Milwaukee, P.O. Box 413,  
Milwaukee, WI 53201}
\author{Eric Gourgoulhon}
\affiliation{
Laboratoire Univers et Th\'eories, UMR 8102 du CNRS,
Observatoire de Paris, Universit\'e Paris Diderot, F-92190 Meudon, France}
\author{Masaru Shibata}
\affiliation{
Yukawa Institute for Theoretical Physics, Kyoto University, Kyoto 606-8502, 
Japan
}
\date{\today}  

\begin{abstract} 

A new method is described for constructing initial data for a binary 
neutron-star (BNS) system in quasi-equilibrium circular orbit.  
Two formulations for non-conformally flat 
data, waveless (WL) and near-zone helically 
symmetric (NHS), are introduced; in each  
formulation, the Einstein-Euler system, written in 3+1 form 
on an asymptotically flat spacelike hypersurface, is exactly 
solved for all metric components, including the spatially 
non-conformally flat potentials, and for irrotational flow.
A numerical method applicable 
to both formulations is explained with an emphasis on 
the imposition of a spatial gauge condition.  
Results are shown for solution sequences of irrotational 
BNS with matter approximated by 
parametrized equations of state 
that use a few segments of polytropic equations of state.  
The binding energy and total angular momentum of solution 
sequences computed within the conformally flat 
--- Isenberg-Wilson-Mathews (IWM) --- formulation are
closer to those of the third post-Newtonian (3PN) 
two point particles up to the closest orbits, for 
the more compact stars, whereas 
sequences resulting from the WL/NHS formulations 
deviate from the 3PN curve even 
more for the sequences with larger compactness.
We think it likely that this correction reflects an overestimation 
in the IWM formulation as well as in the 3PN formula, 
by $\sim 1$ cycle in the gravitational wave phase  
during the last several orbits.  
The work suggests that imposing spatial conformal 
flatness results in an underestimate of the quadrupole 
deformation of the components of binary neutron-star systems 
in the last few orbits prior to merger. 
\end{abstract} 

\maketitle

\section{Introduction}
\label{intro} 

Inspiral to merger of binary neutron stars (BNS) is one of the most
promising sources of ground-based gravitational-wave detectors. 
A fully general relativistic numerical simulation is the unique 
approach to predict the gravitational waveform from the
late inspiral to merger phase.  Such a simulation begins 
with preparing quasi-equilibrium initial data with a close 
orbital separation $\sim 45$--50 km.  

Quasi-equilibrium initial data for binary neutron stars 
introduce two kinds of inaccuracies into inspiral simulations.  
One is due to ignoring the radial component of the velocity of 
orbiting stars, the other to artificial restrictions on  
the geometry of the initial hypersurface  
\cite{miller}.  
A common choice for the geometry of the initial 
hypersurface is a conformally flat three-geometry
\cite{BNSCF,UryuCF}, 
and a similarly restrictive alternative is presented in \cite{NCF}.%
\footnote{For the computation of black hole-neutron star
binary in quasi-equilibrium, see e.g. \cite{BHNS_QE}}  
The former error is reduced by adding radial 
velocity to minimize the oscillation around 
the inspiral orbit, where the radial velocity may be determined 
empirically or calculated from the post-Newtonian formula of 
inspiraling point masses.  
Both errors become negligible if the initial separation of the 
binary is large enough, possibly five orbits or more before the 
merger; but increasing separation increases the cost of computing time, 
and maintaining accuracy in numerical simulations may still be an issue.%
\footnote{For long-term simulation of binary black hole inspirals 
and matching to the post-Newtonian results, see e.g.
\cite{BBH}}

In a previous letter \cite{Uryu:2005vv}, we have reported that 
the inaccuracy of the binary orbit arising from  
spatial conformal flatness can be largely removed 
if one solves the full Einstein equation for all metric 
components, including the non-conformally 
flat part of the spatial metric, on a Cauchy surface 
$\Sigma_t$, 
using the formulation presented in \cite{SG04,BGGN04,SUF04}.  
In this formulation, Einstein's 
equation is written in a 3+1 form and the time derivative of 
the conformal three metric, $\pa_t \tgmabd$, which carries the dynamics 
of the spacetime in our choice of the gauge, is set to zero.
$\tgmabd$ is conformally related to the spatial metric $\gmabd$ in 
each slice $\Sigma_t$ by 
$\gmabd = \psi^4\tgmabd$, with $\psi$ a conformal factor.  
As a result, the field equations 
for the metric components become elliptic equations on an 
initial slice $\Sigma_t$, and they yield an asymptotically 
flat metric.  
We call this approach the waveless formulation (WL).  

We have also experimented with another formulation for 
quasi-equilibrium initial data in which all components of the 
metric are computed; preliminary results were presented in 
\cite{YBRUF06}.  In this approach, 
helical symmetry is imposed in the near zone from 
the center of mass to the radius $\sim \lambda = \pi/\Omega$,
and either the WL formulation is applied outside, or 
the computational domain is truncated at this radius.  
Here, $\Omega$ is the orbital angular velocity and $\lambda$ 
is the wavelength of the dominant, primarily $\ell=m=2$ 
quadrupole, mode of the 
gravitational waves expected to be radiated from the system.  
In this paper, we discuss the near-zone helically symmetric (NHS)
formulation together with the WL formulation, for which numerical 
methods are common.  

A significant difference in the binding energy and total angular momentum 
between the solutions from WL/NHS formulations and those from a conformally 
flat formulation, the Isenberg-Wilson-Mathews (IWM) formulation 
\cite{ISEN78,WM89}, is found and is discussed in the later section.  
In the IWM formulation, the solution sequences -- the plots of 
these quantities as functions of orbital angular velocity -- become 
closer to those of third post-Newtonian (3PN) 
point particles up to the closest orbits when the 
compactness of each star is increased. In contrast, 
sequences obtained from the WL/NHS formulations deviate more 
from the 3PN curve for larger compactness.  

We expect waveless and helically symmetric solutions to 
accurately approximate the outgoing metric in the near zone, 
where the gravitational wave amplitude is small compared to the Coulomb part 
of each metric potential. Results of Ref.~\cite{YBRUF06} and of the present 
paper support this expectation by showing that corresponding WL and NHS 
solutions nearly coincide.

Several groups have developed simulation codes for BNS 
inspirals and merger; stable long-term simulations \cite{BNSmerger,Read:2009yp}, 
magnetized BNS simulations \cite{BNSmagnetized}, and  
black hole-neutron star binary merger simulations \cite{BHNSmerger}
are now feasible.  
As mentioned above, however, accurate modeling of the last several 
orbits of inspiraling binary compact object using quasi-equilibrium 
sequences will be still useful, because the lower computational cost 
allows one to study gravitational wave sources by exploring a wider 
parameter space, varying the mass ratio and the dense matter EOS.  
One of the other applications will be the comparison with the results of simulations, 
which becomes a reliable calibration for both of the numerical solutions.

This paper is organized as follows:
In Sec.\ref{sec:form} we describe the WL/NHS formulations.  
These are essentially identical to those introduced in our previous 
papers \cite{SUF04,BGGN04,YBRUF06}, except for a few 
modifications suitable for coding.  
All equations used in actual coding are written 
in Appendix \ref{secA:eqs} in detail.  
As a model for the EOS of high density matter, 
the parametrized EOS developed in \cite{Read:2009yp,Read:2008iy} 
is used in the computations  
and is briefly introduced in this section.  
In Sec.\ref{sec:comp} the numerical method is discussed, 
with emphasis on the major differences from the previous 
conformally flat code.  
In Sec.\ref{sec:QEsol} we report results from the WL/NHS computation of 
binary systems and of constant rest mass quasi-equilibrium sequences.  

In this paper, spacetime indices are Greek, spatial indices Latin, 
and the metric signature is $-+++$.  For writing the basic equations, 
geometric units with $G=c=1$ are used, while for tabulating the numerical 
solutions, cgs units or other appropriate units are used.

\section{Formulation}
\label{sec:form}

\subsection{3+1 decomposition and gauge conditions}
\label{sec:3plus1}

The spacetime ${\cal M}=\mathbb{R}\times \Sigma$ is foliated 
by a family of spacelike 
hypersurfaces $(\Sigma_t)_{t\in\mathbb{R}}$
parametrized by $t$.
The future-pointing unit normal $n^\alpha$ to the hypersurface $\Sigma_t$ 
is related to the generator $t^\alpha$ of time translations, for which 
$t^\alpha \na_\alpha t = 1$, by 
\beq 
t^\alpha = \alpha n^\alpha + \beta^\alpha.
\eeq
Here  
$\alpha$ is the lapse function and $\beta^\alpha$ the shift vector, 
with $\beta^\alpha n_\alpha=0$.
$n^\alpha$ is related to the gradient of $t$ by
$n_\alpha =-\alpha\na_\alpha t$.  
It is also related to the helical vector $k^\alpha$, 
the generator of time translation in a rotating frame, by 
\beq 
k^\alpha = \alpha n^\alpha + \omega^\alpha, 
\eeq
where a spatial vector 
$\omega^\alpha := \beta^\alpha + \Omega \phi^\alpha$ 
is the rotating shift in the rotating frame, and $\Omega$ 
is a constant angular velocity of the rotating frame.  
The helical vector $k^\alpha$ is not everywhere timelike, 
but it is transverse to the surface $\Sigma_t$, and 
normalized as $k^\alpha \na_t = 1$.  

The spatial metric $\gmabd(t)$ induced on $\Sigma_t$ 
by the spacetime metric $g_{\alpha\beta}$
is equal to the projection tensor orthogonal to $n^\alpha$, 
$\gamma_{\albe} = \gabd+n_\alpha n_\beta$, restricted to $\Sigma_t$.  
We introduce a conformal factor $\psi$, a conformally 
rescaled spatial metric $\tgmabd$, and a flat spatial metric 
$f_{ab}$, with $\gmabd = \psi^4 \tgmabd$, and with the 
conformal factor specified by the condition $\tgamma = f$, 
where $\tgamma$ and $f$ are the determinants of $\tgmabd$ and 
$f_{ab}$.  
In a chart $(t,x^a)$, the metric $\gabd$ has the form
\beq 
ds^2 
= 
-\alpha^2dt^2+\psi^4\tilde \gamma_{ab}(dx^a+\beta^a dt)(dx^b+\beta^b dt) . 
\eeq
Let us denote by $h_{ab}$ and $h^{ab}$ the differences between the conformal metric and the flat one: 
\beq
\tgmabd=f_{ab}+h_{ab}\,,\quad  \tgmabu=f^{ab}+h^{ab}. 
\eeq

The extrinsic curvature of each slice $\Sigma_t$ is defined by 
\beq
\Kabd = -\frac12 \Lie_n\gmabd,  
\label{eq:Kab}
\eeq
where the action of $\Lie_n$ on $\gmabd$ in the above 
definition, and on other spatial tensors hereafter, 
is given by 
\beq
\Lie_n \gmabd := \frac1{\alpha}\pa_t \gmabd - \frac1{\alpha}\Lie_\beta \gmabd;
\eeq
here $\pa_t \gmabd$ is the pullback of $\Lie_t \gamma_\albe$ to $\Sigma_t$,
with  
$\Lie_t$ the Lie derivative along the vector $t^\alpha$ defined on 
$\cal M$, and $\Lie_\beta$ is the Lie derivative along the spatial vector 
$\beta^a$ on $\Sigma_t$.

Einstein's equation is written in the 3+1 form 
\beqn
&&(\Gabd-8\pi\Tabd)n^\alpha n^\beta = 0, 
\label{eq:Hamcon}
\\
&&(\Gabd-8\pi\Tabd)\gmaa n^\beta = 0,  
\label{eq:Momcon}
\\
&&(\Gabd-8\pi\Tabd)\left(\gamma^\albe + \frac12 n^\alpha n^\beta\right) = 0, 
\label{eq:tr}
\\
&&(\Gabd-8\pi\Tabd)\left(\gmaa\gmbb-\frac13\gmabd\gamma^\albe\right) = 0. 
\label{eq:trfree}
\eeqn
These equations are the Hamiltonian and momentum constraints,
the trace of the spatial projection combined 
with the Hamiltonian constraint, 
and the tracefree part of the spatial projection, respectively.  
They are solved for $\psi$, 
$\beta_a$, the combination $\alpha\psi$, and $h_{ab}$.  
For perfect-fluid spacetimes, the stress-energy tensor $\Tabu$ 
is written 
\beq
\Tabu = (\epsilon+p)u^\alpha u^\beta  + p\, \gabu, 
\label{eq:Tab}
\eeq
where $\epsilon$ is the energy density, $p$ the pressure, and $u^\alpha$ 
the 4-velocity of the fluid.  

The above set of equations are solved imposing 
as coordinate conditions the maximal slicing 
condition, 
\beq
K=0,
\label{eq:gauge_maximal}
\eeq 
and the generalized Dirac gauge condition \cite{SUF04,BGGN04,YBRUF06}, 
\beq
\zD_b \tgmabu=\zD_b h^{ab}=0,
\label{eq:gauge_Dirac}
\eeq
where $\zD_a$ is the covariant derivative associated with the flat metric 
$f_{ab}$.  
Concrete forms of Eqs.(\ref{eq:Hamcon})-(\ref{eq:trfree}) 
are presented in Appendix~\ref{secA:eqs}.

\subsection{Waveless and near-zone helically symmetric formulations}
\label{sec:WL_HS}

As a model for binary compact objects in 
general relativity, helically symmetric spacetimes have been 
introduced \cite{BD92D94} and studied by several authors 
\cite{BonazGM97,Friedman:2001pf,GGB02,Klein:2004be,Torre,Bruneton:2006ft,BBS,PSWconsortium}.  
Helically symmetric binary solutions for point-particles in a post-Minkowski 
framework \cite{FU06GU07} analogous to the electromagnetic 
two-body solution \cite{Schild}, 
and for several toy models have been calculated
\cite{YBRUF06}.  

Helically symmetric spacetimes 
do not admit flat asymptotics.  However, it is expected 
that, up to a certain truncation radius where the energy of 
radiation does not dominate the gravitational mass
of the system, solutions have an approximate asymptotic region in which 
gravitational waves are propagating in a curved background.  
Such a solution, however, has not yet been 
calculated successfully in the regime of strong gravity.

Helical symmetry, 
\beq
\Lie_k \gabd = 0, 
\label{eq:helical}
\eeq
implies for the 3-metric and extrinsic curvature 
on a initial hypersurface $\Sigma_t$, 
\beq
\Lie_k \gmabd = 0, \quad  
\Lie_k \Kabd = 0.  
\eeq
Using the relation $k^\alpha = \alpha n^\alpha + \omega^\alpha$,
we have 
\beqn
\Lie_n \gmabd = -\frac1\alpha \Lie_\omega \gmabd, 
\label{eq:Liengmab}
\\ 
\Lie_n \Kabd = -\frac1\alpha \Lie_\omega \Kabd. 
\label{eq:LienKab}
\eeqn
Because $k^\alpha$ is timelike in the fluid, helical 
symmetry for the fluid variables,   
\beq
\Lie_k u^\alpha = 0, \quad  
\Lie_k \epsilon = 0, \quad  
\Lie_k p = 0,  
\label{eq:Liekflu}
\eeq
has the meaning of stationarity for a rotating observer. 

Our formulation for the non-conformally flat data of binary 
compact objects in a quasi-equilibrium quasi-circular orbit 
is based on the helically symmetric formulation.  
We further impose either a waveless condition or near-zone helical 
symmetry in the gauge (\ref{eq:gauge_maximal}) and 
(\ref{eq:gauge_Dirac}).  

\paragraph{Waveless formulation}
As discussed in \cite{SUF04}, 
the condition, 
$\pa_t\tgmabu=\Od(r^{-3})$, is sufficient to enforce 
Coulomb-type fall off 
in the asymptotics.  For our waveless formulation 
in this paper, we impose the stronger condition 
\beq
\pa_t\tgmabd=0,
\label{eq:WL_cond}
\eeq
which amounts to writing the extrinsic curvature as 
\beqn
\Kabd &=& 
\frac1{2\alpha} \Lie_\beta \gmabd -\frac1{2\alpha}\gmabd
\left(\frac{\tgamma}{\gamma}\right)^{\frac13}
\pa_t\left(\frac{\gamma}{\tgamma}\right)^{\frac13}
\nonumber\\
 &=&
 \frac1{2\alpha} \Lie_\beta \gmabd +\frac1{2\alpha}\gmabd\,
\Omega\Lie_{\phi}\ln\psi^{4}, 
\label{eq:Kab_WL}
\eeqn
where helical symmetry is used to get the second equality.  
Only the first term on the r.h.s. remains 
in the maximal slicing condition. 
Because the trace of Eq.~(\ref{eq:Kab_WL}) has the same 
form for $K=0$ as the trace of the original equation (\ref{eq:Kab}), 
the waveless condition (\ref{eq:WL_cond}) does not 
affect the maximal slicing condition.  Note that the second term 
of the r.h.s. of Eq.~(\ref{eq:Kab_WL}) does not appear 
in the tracefree part of $K_{ab}$; in other words, the time 
derivative of the conformal factor $\psi$ does not appear 
in the initial value formulation in this slicing.  
The other time derivatives are given by the 
helical symmetry conditions, Eqs.(\ref{eq:LienKab}) 
and (\ref{eq:Liekflu}). 

\paragraph{Near-zone helically symmetric formulation}

Near-zone helical symmetry means that we impose helically 
symmetric conditions (\ref{eq:Liengmab})-(\ref{eq:Liekflu})
in the region from the center of the source to about 
one wavelength of the $\ell=m=2$ mode of the gravity wave, 
$r \alt \lambda:=\pi/\Omega$; we then 
either truncate the domain of numerical computation 
at this radius or 
use the waveless formulation outside.  
The latter implies for $\Kabd$ the condition 
\beq
\!\!\!
\Kabd = \left\{
\begin{array}{ll}
\displaystyle
\frac1{2\alpha}\Lie_\omega\gmabd 
& 
\mbox{for} \ r < a \lambda,  \\ \\
\displaystyle
\frac1{2\alpha}\Lie_\beta\gmabd
+\frac1{2\alpha}\gmabd\,
\Omega\Lie_{\phi}\ln \psi^{4}
&
\mbox{for} \ r \geq a \lambda,   
\end{array}
\!\!\!\!\right. 
\label{eq:helicutoff}
\eeq
where the constant $a$, the coordinate radius of the 
helically symmetric zone in units of $\lambda=\pi/\Omega$, 
is restricted to $a\alt 1.5$. Without this restriction, iterations 
fail to converge to a binary solution.  
In the near-zone-helical + outside-waveless formulation, 
all metric components, including those of 
the spatial metric, have Coulomb-type fall off.  
We have compared the NHS solution 
to the WL solution in our previous paper \cite{YBRUF06}
and confirmed that the difference in the non-conformal flat part 
of the spatial metric is about 1\% for the BNS 
of $\compa \sim 0.17$, where $\compa$ is the compactness, 
the ratio of the gravitational mass to the circumferential radius
of a spherical star having the same rest mass as each 
component star of the binary.

\subsection{Formulation for the irrotational flow}
\label{sec:matter}

The late stage of BNS inspiral  
is modeled by a constant rest mass sequence of 
quasi-equilibrium solutions 
with negligible spins and magnetic fields, a description
appropriate to a binary of old pulsars with spin periods 
longer than $100$ ms.  Since the viscosity of the high density matter 
is expected to be negligible, a neutron star in a binary system 
is not spun up by the tidal torque during the inspirals.  
Hence, the flow field remains approximately irrotational, and 
each neutron star is modeled by an irrotational 
perfect fluid \cite{KBC92}.

The equation of motion,
$ 
\na_\beta \Tba=0, 
$ 
for a perfect fluid has the form, 
\beqn
{ \na_\beta \Tba}
&=& 
\rho\biggl[\,{ u^\beta \na_\beta(hu_\alpha) + \na_\alpha h}
\,\biggr]
\nonumber \\
\,&+&\, hu_\alpha { \na_\beta(\rho u^\beta)}
\,-\, \rho T{ \na_\alpha s} \,=\, 0 ,  
\eeqn
where $s$ is the entropy per baryon mass, 
$h$ is the relativistic enthalpy per baryon mass
$h:=(\epsilon+p)/\rho$, and 
local thermodynamic equilibrium
$dh = Tds + dp/\rho$ is assumed.  
We assume constant entropy per baryon ($s = {\rm const}$)  
everywhere inside the neutron star, together with a one-parameter 
EOS, 
\beq
p = p(\rho).  
\label{eq:one_para_EOS}
\eeq
The form 
\beq
u^\beta \na_\beta(hu_\alpha) + \na_\alpha h\,=\,0 
\label{eq:Euler_eq}
\eeq
of the relativistic Euler equation then follows from local conservation 
of baryon mass, 
\beq
\na_\alpha (\rho u^\alpha)
\,=\,0. 
\label{eq:masscon_eq}
\eeq
Written in terms of the Lie derivative along $u^\alpha$,
these last equations have the form 
\beqn
&\frac{1}{\sqrt{-g}}\Lie_u (\rho\sqrt{-g})
\,=\,0, &
\label{eq:masscon_eqL}
\\
&\Lie_u (hu_\alpha) + \na_\alpha h\,=\,0. &
\label{eq:Euler_eqL}
\eeqn

A state is stationary state in the rotating frame if it 
is helically symmetric, if each physical field is Lie 
derived by the helical vector field $k^\alpha$, as 
in Eq.~(\ref{eq:Liekflu}), or 
\beq
\Lie_{k}(\rho u^t \sqrt{-g})=0, \ \ \mbox{and}\ \ 
\Lie_{k} (h u_\alpha)=0, 
\label{eq:helical_fluid}
\eeq 
where $u^t$ is the scalar $u^\alpha \na_\alpha t$.

The relativistic Euler equation (\ref{eq:Euler_eq}) 
can be rewritten as 
\beq
u^\beta\omega_{\beta\alpha} = 0, 
\eeq
where 
\beq
\omega_{\beta\alpha}:=\na_\beta (hu_\alpha)- \na_\alpha(hu_\beta)  
\eeq
is the relativistic vorticity tensor.
This implies that, for irrotational flow, $hu_\alpha$ has 
a potential $\Phi$, 
\beq
h u_\alpha = \na_\alpha \Phi, 
\label{eq:vpot}
\eeq
and hence the relativistic Euler equation has a first integral.  
With a spatial velocity $v^\alpha$ in the 
rotating frame defined by 
\beq
 u^\alpha = u^t (k^\alpha+ v^\alpha),
\label{eq:udecomp}
\eeq
where $v^\alpha n_\alpha=0$, Eq.~(\ref{eq:Euler_eqL}) becomes,  
\beqn
\Lie_u (hu_\alpha) + \na_\alpha h
&=&
u^t\left[\Lie_{k+v} (hu_\alpha) + \na_\alpha \frac{h}{u^t}\right]
\nonumber\\
&=&
u^t\na_\alpha \left(\Lie_{v} \Phi + \frac{h}{u^t}\right)
\,=\,0; 
\eeqn
therefore the first integral is 
\beq
\Lie_{v} \Phi \,+\, \frac{h}{u^t} \,=\, {\cal E}, 
\label{eq:firstint}
\eeq
where ${\cal E}$ is a constant
\footnote{
Cartan identity $k^\beta \omega_{\beta\alpha} 
= \Lie_k(hu_\alpha)-\na_\alpha(hu_\beta k^\beta) $
implies, for the helically symmetric irrotational flow 
satisfying $\Lie_k (hu_\alpha) = 0$ and 
$\omega_{\beta\alpha} =0$, 
a relation, $hu_\alpha k^\alpha = {\rm constant}$, equivalent to 
Eq.~(\ref{eq:firstint}). 
}.  
Note that Eqs.~(\ref{eq:helical_fluid})
and (\ref{eq:vpot}) imply a flow with $\Lie_k\Phi = constant$.  
Such a flow is both irrotational and helically symmetric with 
the shape of the star fixed in the rotating frame.  
Solutions describing irrotational binaries in Newtonian and 
post-Newtonian gravity are found in \cite{irbns_Newtonian}, and 
details of the formulation for helically symmetric 
irrotational flow are given in \cite{BonazGM97,irbns_formulation}.

There are three fluid variables and two parameters 
to be determined in the above formulation.  The fluid 
variables are a thermodynamic variable, 
the velocity potential $\Phi$, and the time component of 
the 4-velocity $u^t$; and these are calculated from 
the first integral (\ref{eq:firstint}), the rest-mass conservation 
equation (\ref{eq:masscon_eqL}), and the normalization 
of the 4-velocity, $u_\alpha u^\alpha = -1$.  A concrete 
form of these equations are presented in Appendix 
\ref{sec:irrotfluid}.  
For the independent thermodynamic variable, we choose 
$q:=p/\rho$, and other thermodynamic variables are 
determined from the thermodynamic relations and 
the one-parameter EOS, which are briefly 
explained in the next section.  
The number of fluid variables and parameters are augmented 
in the numerical computation, which is mentioned in 
Appendix \ref{secA:numeiter} (or see \cite{Huang:2008vp}).

\subsection{Parametrized equations of state}
\label{sec:pEOS}

Recently, a parametrization for the EOS of nuclear matter 
has been studied, and it is shown that 
a parametrized EOS with three polytropic intervals 
approximates with fair accuracy a variety of current candidate
EOS, over a  
range of densities that extends from the 
inner crust to the maximum neutron-star density \cite{Read:2008iy}. 
Two of these intervals and three parameters cover densities 
below the central density of a $1.4 M_\odot$ neutron star, 
and waveforms from binary inspiral can be used to constrain 
this three-dimensional subspace of the parameter space \cite{Read:2009yp}.  
This parametrized EOS is used in our models for 
BNS data. 

\subsubsection{Construction of piecewise polytropic EOS}

In presenting these piecewise polytropes, it is helpful 
to introduce a relativistic Emden function $q$ by
\beq  
 q:=p/\rho
\eeq
and to write the remaining thermodynamic variables in terms of $q$.
For an isentropic flow with $s=0$, the local first law of thermodynamic 
equilibrium, $dh = Tds + \frac1{\rho}dp$, takes the form
\beq 
dh = \frac1{\rho}dp, 
\eeq
where $h$ is the enthalpy per baryon mass.  

A piecewise polytropic EOS is given by  
\beq 
p = K_i \rho^{\Gamma_i}, 
\eeq
in the intervals $\rho \in [\rho_{i-1}, \rho_i)$, 
$i = 1, \cdots, N$, with $\rho_0=0$ and 
$\rho_N=\infty$.  In this section the subscript 
$i$ denotes the $i$th interval, associated with a set of 
constants $\{\Gamma_i,K_i\}$ with $i = 1, \cdots, N$,
and labels the value of quantities 
at the higher density side of each interval, 
$[\rho_{i-1}, \rho_i)$.  Because we consider only 
continuous EOS, $p_i, h_i, \epsilon_i$ and $q_i$ 
are the values of each of these quantities at density $\rho_i$.

The constant indices $\Gamma_i$ are $N$ model parameters, 
and values of one thermodynamic variable at interfaces 
comprise a set of $N-1$ model parameters.  
A requirement that the pressure at the interface is 
continuous, 
\beqn
K_{i}\rho_i^{\Gamma_{i}} &=& K_{i+1}\rho_i^{\Gamma_{i+1}}, 
\label{eq:adconi}
\eeqn
uniquely specifies values of $K_i$ up to one free parameter, 
one of $K_i$ of a specific $i$th interval, 
which is usually specified by prescribing the value of 
pressure $p_i$ at the corresponding interface density $\rho_i$.  
Therefore we have $2N$ parameters for a parametrized EOS 
with $N$ intervals.

To compute other thermodynamic 
quantities from $q$, we use the following relations, valid in the 
$i$th interval, $q \in [q_{i-1},q_{i})$: 
\beqn
\rho
&=& K_i^{\frac{-1}{\Gamma_i-1}}q^{\frac1{\Gamma_i-1}},
\label{eq:rhoi}
\\ \nonumber \\
p
&=& K_i^{\frac{-1}{\Gamma_i-1}} q^{\frac{\Gamma_i}{\Gamma_i-1}},
\label{eq:prei}
\\ \nonumber \\
h-h_{i-1}
&=& \frac{\Gamma_i}{\Gamma_i-1} (q-q_{i-1}),
\label{eq:hi}
\\ \nonumber \\
\epsilon &=& \rho h - p,
\label{eq:epsi}
\eeqn
where Eq.~(\ref{eq:hi}) is obtained by integrating the relation 
\beq
dh = \frac1{\rho}dp = \frac{\Gamma_i}{\Gamma_i-1}d q 
\eeq
in the $i$th interval $q\in[q_{i-1},q_i]$.  
Here, 
\beq
h_i = h_0 
+ \sum_{j=1}^{i}\frac{\Gamma_j}{\Gamma_j-1}(q_j-q_{j-1}), 
\eeq
with $h_0 = 1$ and $q_0=0$.

\subsubsection{Choice for the parameters}

In the latter sections, we present the results of 
quasi-equilibrium BNS solutions 
calculated using two types of parametrized EOS.  
The first EOS contains one free parameter, 
which is used to estimate the accuracy of the measurement 
of the EOS parameter, and the neutron-star radius, 
by gravitational-wave observations of the inspirals of BNS 
\cite{Read:2009yp}.  The second EOS is a four-parameter fit 
to the candidates of neutron-star EOS.  Those candidate EOS 
are tabulated nuclear EOS, and the parametrized EOS with 
four parameters approximates each candidate within the rms residual 
typically in the order of $\sim 0.1\%$, and $\sim 4.3\%$ for the worst case 
\cite{Read:2008iy}.  

The parametrized EOS with one parameter 
uses two polytropic intervals.  The lower 
density interval approximates the known subnuclear density EOS, 
the fixed crust EOS, around $0.1 \rho_{\rm nuc}\sim \rho_{\rm nuc}$ 
by setting $(\Gamma_0,K_0) = 
(1.35692,3.59389\times 10^{13})$.  Here, 
$\rho_{\rm nuc}$ is the nuclear saturation density, and 
the constant $K_0$ is in cgs units which give the pressure 
$p$ in dyn/cm$^2$.  
For the second polytropic interval at the higher density side, 
the adiabatic index is set $\Gamma_1=3$.  Then, the pressure 
$p_1$ at the density $\rho_1 = 10^{14.7}$ g/cm$^3$ is 
chosen as a parameter, and the dividing density at 
the fixed crust and the next polytropic piece $\rho_0$ 
is determined as the intersection of the two intervals.  
Further details are found in \cite{Read:2009yp}.  

The four-parameter fit uses the same crust EOS as above, 
and three other polytropic intervals.  
The adiabatic indices of higher polytropic intervals 
$\{\Gamma_1, \Gamma_2, \Gamma_3\}$, and the pressure 
$p_1$ at the interface between $i=1$ and $2$ are 
chosen as fitting parameters, while 
the dividing density $\rho_0$ is evaluated in the same way 
as above, and other dividing densities are fixed as 
$\rho_1 = 10^{14.7}$ g/cm$^3$ and $\rho_2 = 10^{15}$ g/cm$^3$.  The EOS parameters 
and corresponding data for the spherical solutions are 
summarized in the later section \ref{sec:QEsol}.
Further details for the four-parameter fit 
are found in \cite{Read:2008iy}.

\section{Computation}
\label{sec:comp}

A system of elliptic equations and algebraic relations 
are solved applying a self-consistent field iteration scheme 
\cite{OM68}.  
Recently, the convergence of such scheme for Newtonian 
barotropic stars has been mathematically analyzed in \cite{Price:2009nv}.

The WL/NHS code for the irrotational BNS 
presented in this paper is developed on top of the former 
BNS code in which the IWM formulation is used 
\cite{USE00}.  
Another version of the WL/NHS code, based on the triaxially 
deformed rotating neutron-star code described in \cite{Huang:2008vp}, 
has been developed, and its results are presented elsewhere.  
The numerical method used in these codes is briefly 
repeated in Appendix \ref{secA:nume}.

\subsection{Imposition of Dirac gauge}

The primary difference between the WL/NHS code and an IWM code
is the computation of the non-conformally 
flat part of the spatial metric $\tgmabd = f_{ab}+h_{ab}$.  
The conformal spatial metric $\tgmabd$ has to satisfy two 
conditions, $\tgamma=f$ and $\zD_b \tgmabu=0$, which 
turn out not to be automatically satisfied when the 
spatial tracefree part of Einstein's equation 
(\ref{eq:trfree}) or its concrete form in the code, either 
Eq.~(\ref{eq:WLtrfree}) or (\ref{eq:NHStrfree}), 
is solved for $h_{ab}$.  
To impose these conditions on $\tgmabd$ accurately, 
we first make a gauge transformation of $h_{ab}$ 
to satisfy $\zD_b h^{ab}=0$, and we then correct  
the conformal factor to enforce the 
relation $\gamma = \psi^{12} f$ at each 
iteration cycle.  Note that these 
two conditions are not explicitly imposed in Eq.~(\ref{eq:WLtrfree}) or 
(\ref{eq:NHStrfree}), and they are violated mainly 
due to the numerical error of finite differencing.  

The gauge vector is calculated numerically 
by the following procedure: 
a perturbation of the spatial metric $\dl\gmabd$
\beq
\dl \gmabd \rightarrow \dl \gmabd - \zD_a \xi_b - \zD_b\xi_a, 
\eeq
implies, to the same order, that the conformally 
rescaled metric with $\tgamma = f$ satisfies
\beq
\dl \tgmabd \rightarrow \dl \tgmabd - \zD_a \xi_b - \zD_b\xi_a
+\frac23 f_{ab}\zD_c \xi^c. 
\eeq
We adjust $h_{ab}$ to this order to satisfy the Dirac gauge condition; 
namely, writing   
\beq
h'_{ab} = h_{ab} - \zD_a \xi_b - \zD_b\xi_a
+\frac23 f_{ab}\zD_c \xi^c, 
\label{eq:Dgaugetransf}
\eeq
we let $h'_{ab}$ satisfy the Dirac gauge condition to  
linear order in $h_{ab}$, $\zD^b h'_{ab}=0$, 
which leads to 
\beq
\zLap \xi_a + \frac13 \zD_a \zD^b \xi_b = \zD^b h_{ab}.
\label{eq:DgaugeEq}
\eeq

This equation is solved by introducing the decomposition 
\beq
 \xi_a = G_a - \frac14 \zD_a B,
\label{eq:DgaugeBY}
\eeq
which results in a set of elliptic equations,
\beqn
\zLap G_a = \zD^b h_{ab},
\ \mbox{ and }\ 
\zLap B = \zD^a G_a.
\label{eq:Dgauge_ellip}
\eeqn
These equations (\ref{eq:Dgauge_ellip}) are solved using the same 
Poisson solver described in Appendix \ref{secA:nume}, 
and a solution is substituted in Eq.~(\ref{eq:DgaugeBY}) 
and then in Eq.~(\ref{eq:Dgaugetransf}).  
In the r.h.s.~of Eq.~(\ref{eq:Dgaugetransf}), $h_{ab}$ is 
calculated from the tracefree part of Einstein's equation, 
either Eq.~(\ref{eq:WLtrfree}) or (\ref{eq:NHStrfree}), 
and it is replaced by $h'_{ab}$, which 
satisfies the Dirac gauge condition more accurately.  
We have also experimented with a transformation 
of the contravariant components of $h^{ab}$ analogous to 
Eq.~(\ref{eq:Dgaugetransf}), and let 
$\zD_b h^{ab}=0$ be satisfied; however, the results 
did not change.  

After the above gauge transformation, 
the condition $\tgamma = f$ is imposed by
adjusting the conformal factor $\psi$ to 
\beq
\psi' = \psi\left(\frac{\tgamma'}{f}\right)^{\frac{1}{12}}, 
\eeq
where $\tgamma'$ is the determinant of 
$\tgmabd' = f_{ab} + h'_{ab}$.  Note that, to impose 
$\tgamma = f$, we do not change the value of $h'_{ab}$.  
These two corrections to $h_{ab}$ and $\psi$ 
are made once per iteration 

The other parts of the method of computation, 
including the iteration 
scheme, are common to our previous codes 
\cite{USE00,Huang:2008vp,Tsokaros:2007rb},
which are briefly reviewed in Appendix \ref{secA:nume}.

\subsection{Coordinate and grid parameters}
\label{sec:coord}

The WL/NHS code uses two coordinate patches: a spherical patch,
called the central coordinate system, 
on which the metric components are 
calculated, 
and a surface-fitted spherical coordinate patch on which 
the fluid variables are computed.  
The origin of the central coordinates $(r,\theta,\phi)$ 
is the mass center of the binary system, and that of the 
surface-fitted coordinates $(\hat{r}_f,\theta_f,\phi_f)$ 
is the geometric center of the component star, where 
$\hat{r}_f$ is related to the radial coordinate $r_f$ by 
$\hat{r}_f = r_f /R(\theta_f,\phi_f)$ and $R(\theta_f,\phi_f)$ 
is the surface of the star.  We match the radial coordinate 
lines at $(\theta,\phi)=(\pi/2,0)$ of the central coordinates, 
and that of the surface-fitted coordinates 
$(\theta_f,\phi_f)=(\pi/2,0)$, and set the $\theta=0$
and $\theta_f=0$ lines to be parallel.  
Only the octant of the 
whole space for the central coordinate is solved, 
while a quarter for the surface-fitted coordinate. 
The spherical coordinates correspond to the Cartesian 
coordinates in the usual way; the $(\theta,\phi)=(\pi/2,0)$ line 
to the (positive) $x$-axis, the $(\pi/2,\pi/2)$ line to 
the $y$-axis, and the $\theta=0$ line to the $z$-axis.  

The accuracy of the numerical solutions depends on 
the resolution of the finite differencing determined by 
the grid spacings $(\Dl r,\Dl \theta,\Dl \phi)$, 
and the order of the truncation of 
multipole expansion $\ell_{\rm max}$.  
The latter is constrained by the resolution since the multipoles 
involved in the Green's function, which oscillates 
rapidly for the larger $\ell$, should be resolved 
on the grids.  
The radial grid spacing $\Dl r$ of the central coordinates 
is equidistant for $r\in[0,r_c]$, and increases in geometric 
progression for $r\in[r_c,r_b]$.  The grid spacings of the 
other coordinates are equidistant.  
For further details, see \cite{USE00,Huang:2008vp,Tsokaros:2007rb}.  
For the grid parameters, 
we choose values listed in Table \ref{tab:reso}.  
Typically, 1 cycle of iteration takes about 70 s for this 
grid setup using a single core of Intel Xeon CPU X5450 
with 3.00GHz clock.  

\begin{table}
\begin{tabular}{lll}
\hline
$N_{r}$ &:& Number of intervals $\Dl r_i$ in $r \in[0,r_{b}]$ (CC). \\
$n_{r}$ &:& Number of intervals $\Dl r_i$ in $r \in[0,r_{c}]$ (CC). \\
$N_{\theta}$ &:& Number of intervals $\Dl \theta_i$ in $\theta\in[0,\pi/2]$ (CC). \\
$N_{\phi}$ &:& Number of intervals $\Dl \phi_i$ in $\phi\in[0,\pi/2]$ (CC). \\
$N_{\hat{r}}^f$ &:& Number of intervals $\Dl \hat{r}_{i}$ in 
$\hat{r}_f\in[0,1]$ (SFC). \\
$N_{\theta}^f$ &:& Number of intervals $\Dl \theta_i$ in $\theta_f\in[0,\pi/2]$ (SFC). \\
$N_{\phi}^f$ &:& Number of intervals $\Dl \phi_i$ in $\phi_f\in[0,\pi]$ (SFC). \\
\hline
\end{tabular}  
\caption{Summary of grid parameters. (CC) stands for the 
central coordinates, and (SFC) for the surface-fitted 
coordinates.}
\label{tab:grid}
\end{table}

\begin{table}
\begin{tabular}{ccccccccccc}
\hline
$r_b$&$r_c$&$N_r$&$n_r$&$N_\theta$&$N_\phi$&$l_{\rm max}$&
$N_{\hat{r}}^f$&$N_{\theta}^f$&$N_{\phi}^f$&$l^f_{\rm max}$ \\
\hline
$10^4 R_0$& 5$R_0$& 250& 160& 64& 64& 40& 32& 32& 24& 8 \\
\hline
\end{tabular}
\caption{Coordinate parameters, and the number of grid points 
used in this paper.  $R_0$ is the geometrical radius 
of the neutron star along the $(\theta_f,\phi_f)=(\pi/2,0)$ line.  
$l_{\rm max}$ and $l^f_{\rm max}$ are the highest multipoles included 
in the Legendre expansion in the central and surface-fitted coordinates, 
respectively.}
\label{tab:reso}
\end{table}

\begin{table*}
\begin{tabular}{ccccccccccc}
\hline
Model & $\log(\rho_0)$ & $\log(p_1)$ & $\Gamma_1$ & $\Gamma_2$ & $\Gamma_3$&
$M_1\,[M_\odot]$ & $M_0\,[M_\odot]$ & $R$ [km] & $\compa$ & $\log(\rho_c)$ \\
\hline
2H    & 13.847 & 34.90& 3& 3& 3& 1.35& 1.4549& 15.224& 0.13097 & 14.573 \\
HB    & 14.151 & 34.40& 3& 3& 3& 1.35& 1.4927& 11.606& 0.17181 & 14.918 \\
2B    & 14.334 & 34.10& 3& 3& 3& 1.35& 1.5251& 9.7268& 0.20500 & 15.141 \\
SLy   & 14.165 & 34.384& 3.005& 2.988& 2.851& 1.35& 1.4947& 11.469& 0.17385 & 14.934 \\
APR1  & 14.294 & 33.943& 2.442& 3.256& 2.908& 1.35& 1.5388& 9.1385& 0.21819 & 15.221 \\
FPS   & 14.220 & 34.283& 2.985& 2.863& 2.600& 1.35& 1.5055& 10.702& 0.18631 & 15.038 \\
BGN1H1& 14.110 & 34.623& 3.258& 1.472& 2.464& 1.35& 1.4789& 12.626& 0.15792 & 14.912 \\
ALF3  & 14.188 & 34.283& 2.883& 2.653& 1.952& 1.35& 1.5069& 10.350& 0.19264 & 15.150 \\
\hline
\end{tabular}
\caption{
Parameters of each EOS and properties of the spherical neutron-star 
model based on that EOS and having gravitational mass $M_1=1.35 M_\odot$.  
The pressure $p_1$ [dyn/cm$^2$] is the value at the dividing density 
$\rho_1 = 10^{14.7}$ g/cm$^3$, and values of 
$\log(p_1)$ and $\{\Gamma_1,\Gamma_2,\Gamma_3\}$ are taken from 
Table I of \cite{Read:2009yp} and Table III of \cite{Read:2008iy}.  
The parameters to fit the crust EOS are chosen as  
$(\Gamma_0,K_0) = (1.35692,3.59389\times 10^{13})$
where $K_0$ is in cgs units, and the dividing 
density $\rho_0$ used to model the transition from 
the crust to the nuclear matter is tabulated in the $\log$
of $\rho_0$ [g/cm$^3$].  
In the following calculations for BNS, 
a spherical solution of each EOS with gravitational mass 
$M_1 =1.35 M_\odot$ is used as a reference, 
whose rest mass $M_0$ in solar mass units, circumferential 
radius $R$ in km, compactness $\compa$ in the geometric unit $G=c=1$, 
and $\log$ of the central density $\rho_c$ in g/cm$^3$ 
are tabulated.  
}
\label{tab:EOS}
\end{table*}

\section{Quasi-equilibrium solutions}
\label{sec:QEsol}

%
%
\begin{figure}
\begin{tabular}{cc}
\begin{minipage}{.5\hsize}
\begin{center}
\includegraphics[height=40mm]{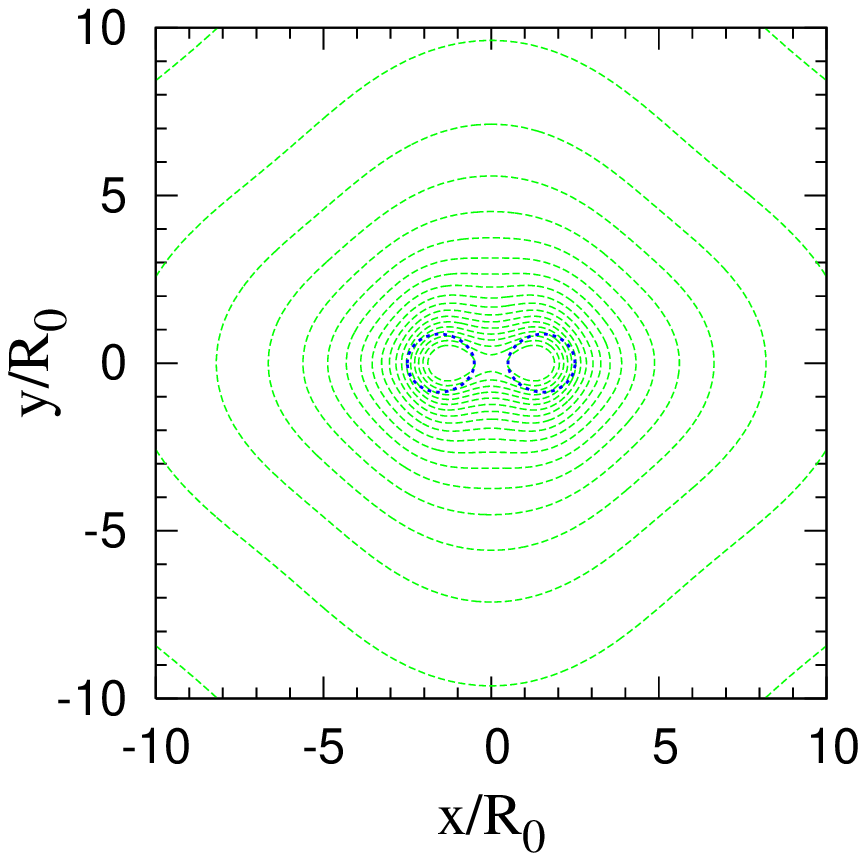}
\includegraphics[height=38mm]{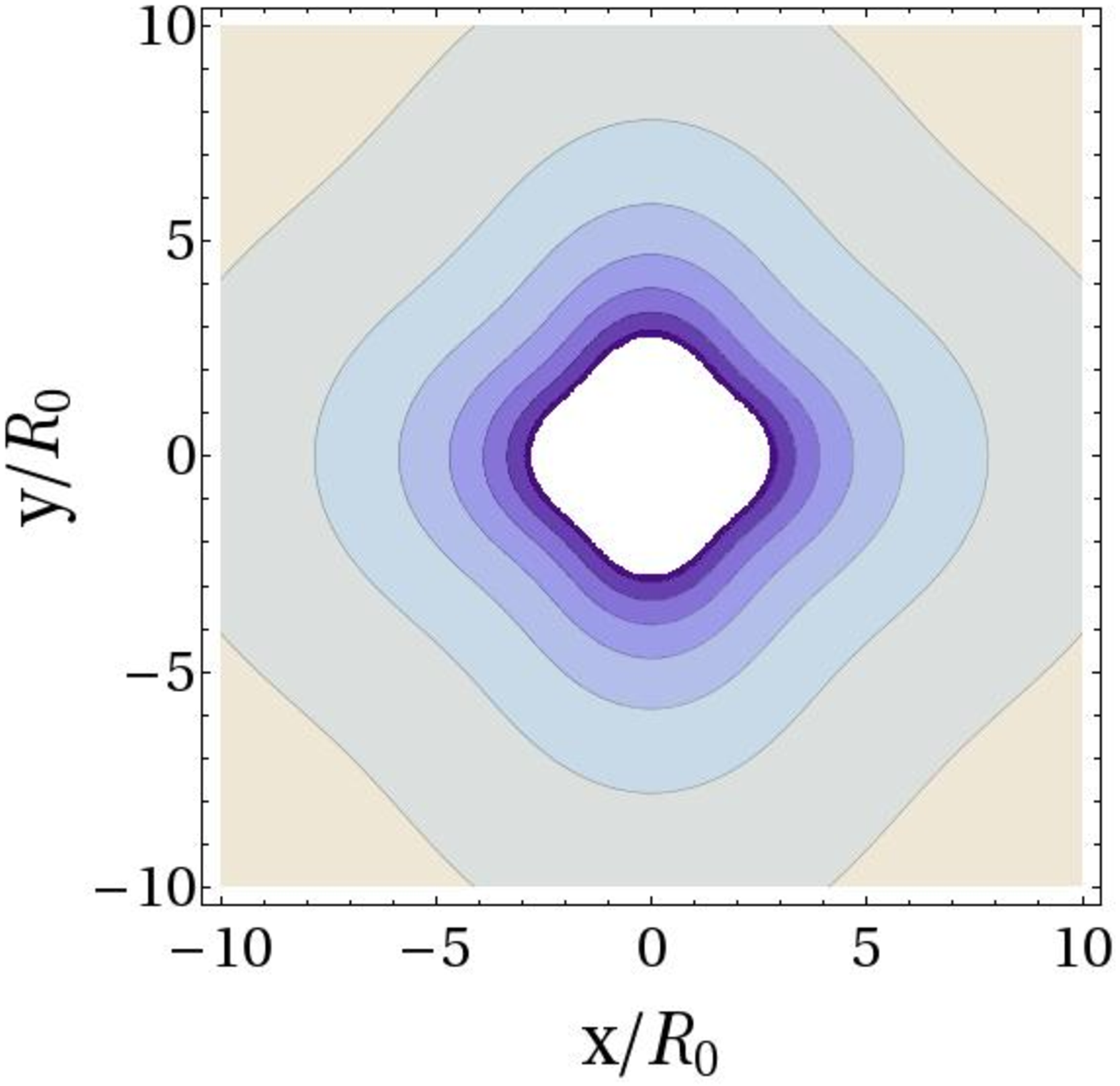}
\end{center}
\end{minipage} 
&
\begin{minipage}{.5\hsize}
\begin{center}
\includegraphics[height=40mm]{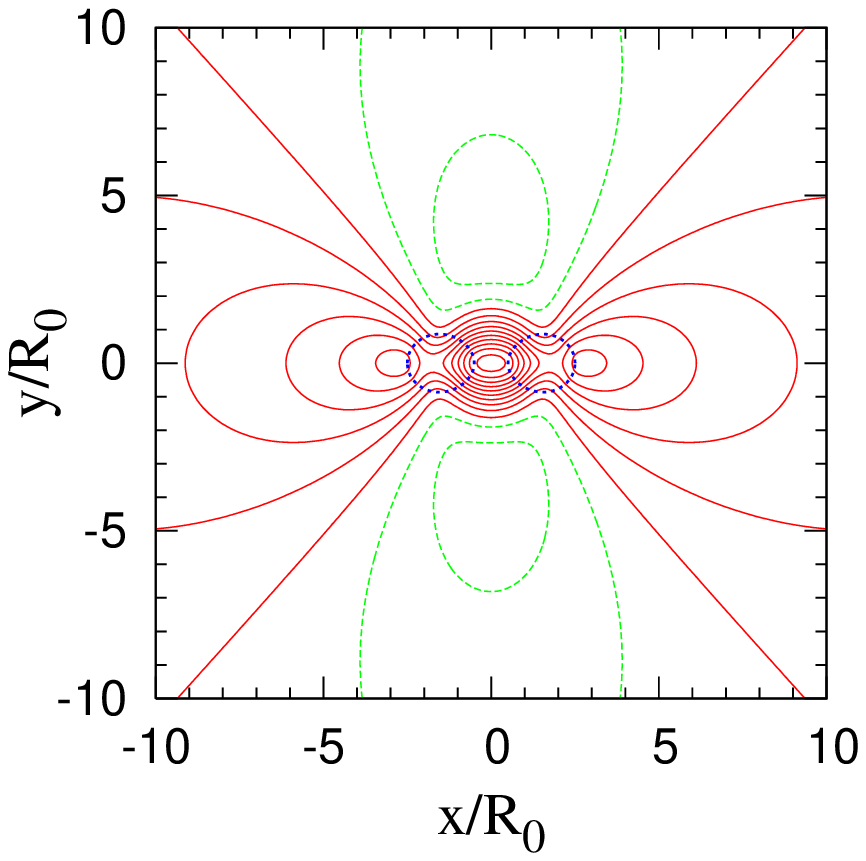}
\includegraphics[height=38mm]{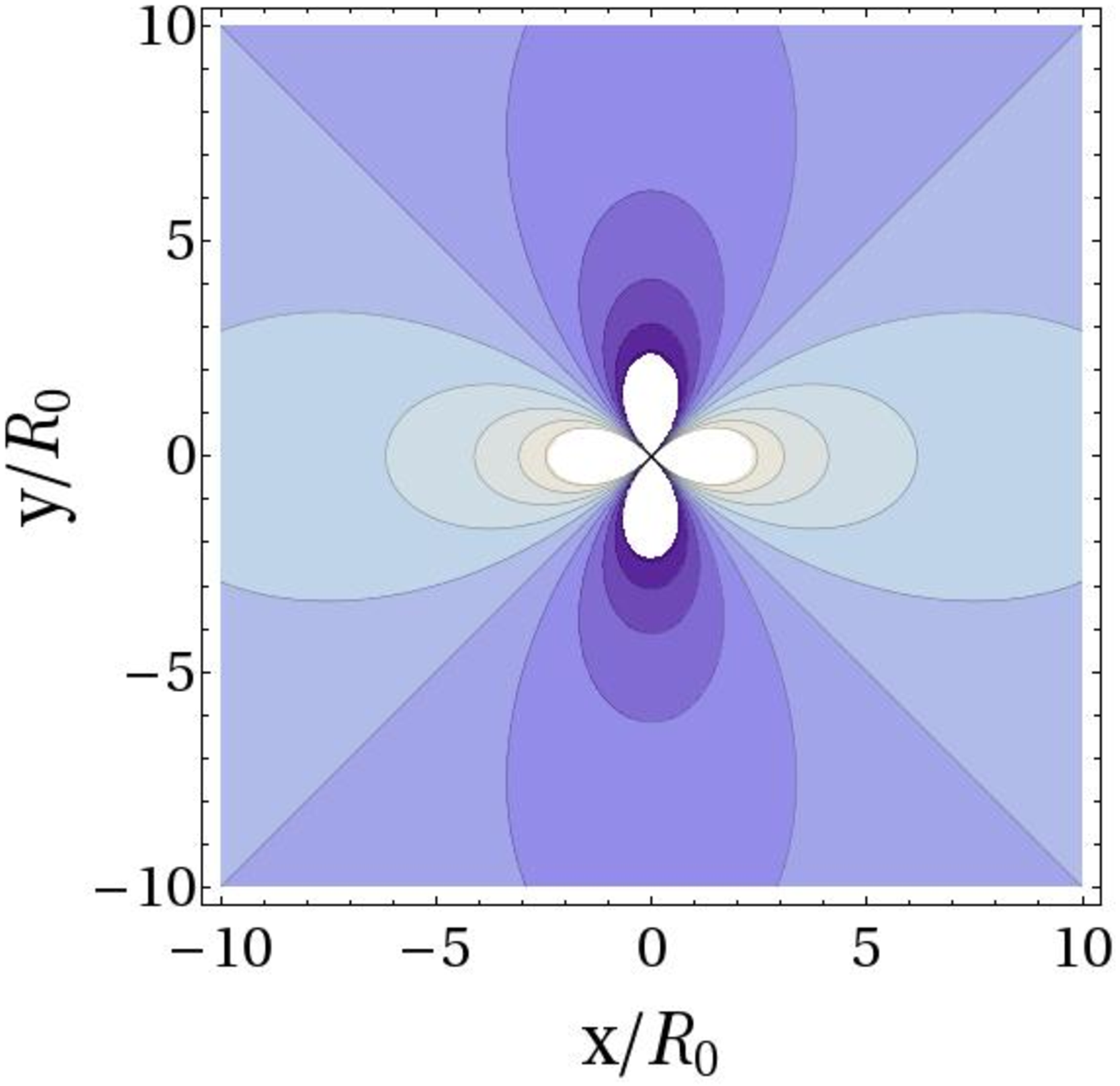}
\end{center}
\end{minipage} 
\end{tabular} 
\caption{Contours of $(h_{xx}-h_{yy})/2$ (left panels) 
and $h_{zz}$ (right panels) in the $xy$-plane.  
Top panels are those of a WL solution with 
the EOS parameter HB and the orbital radius $d/R_0$ = 1.5, 
where $R_0$ is the coordinate radius (a half of the diameter) 
of the neutron star along the $x$-axis.  
Contours are drawn every 0.001 step, where the solid (dashed) 
contours in the top panels corresponds to positive (negative)
values of $h_{ij}$.  Thick dotted circles are the surface of 
neutron stars.  
Bottom panels are the contours of the 2PN 
asymptotic formula (\ref{eq:2PN_hij_asymp}) 
calculated for the two point masses assuming the same 
coordinate separation $d/R_0=1.5$, and mass 
$M= 1.35 M_\odot \times 2$.  Contours are also 
drawn every 0.001 step.  
In the left top and bottom panels, the outermost contour 
(interrupted by the boundary of the figure) corresponds 
to $-0.002$.  In the bottom two panels, contours out of range 
($< -0.01$ for the left panel, and $\gtrless \pm 0.005$ for the right) 
are truncated.}
\label{fig:hab_contour}
\end{figure}
\begin{figure*}
\begin{tabular}{cc}
\begin{minipage}{.5\hsize}
\begin{center}
\includegraphics[height=35mm]{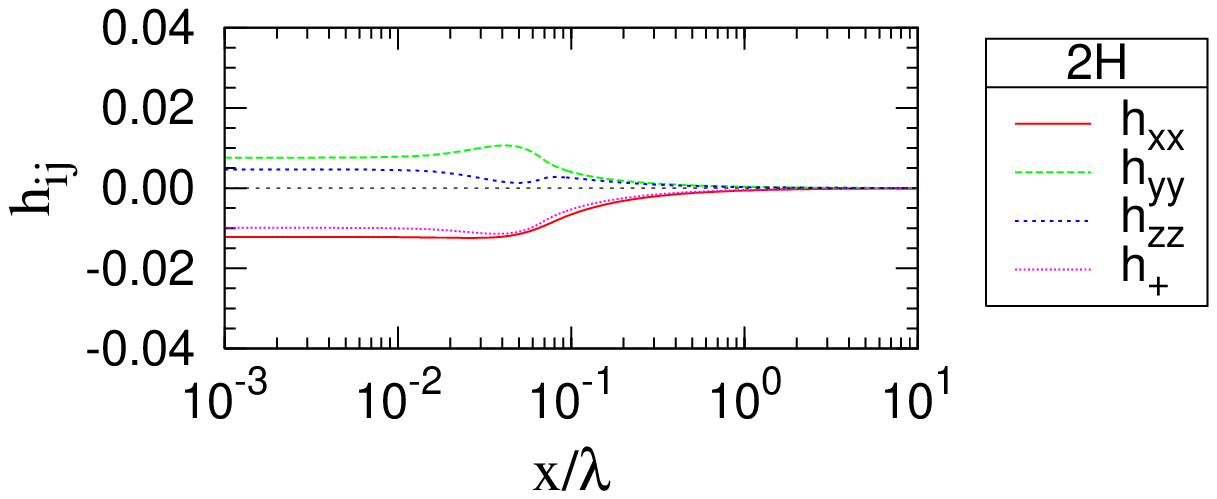}
\includegraphics[height=35mm]{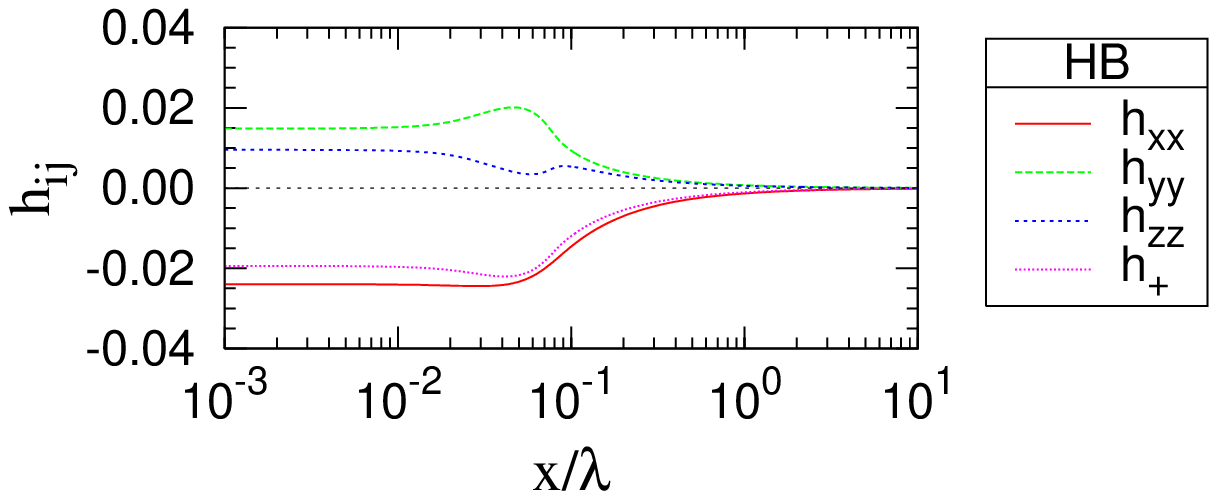}
\includegraphics[height=35mm]{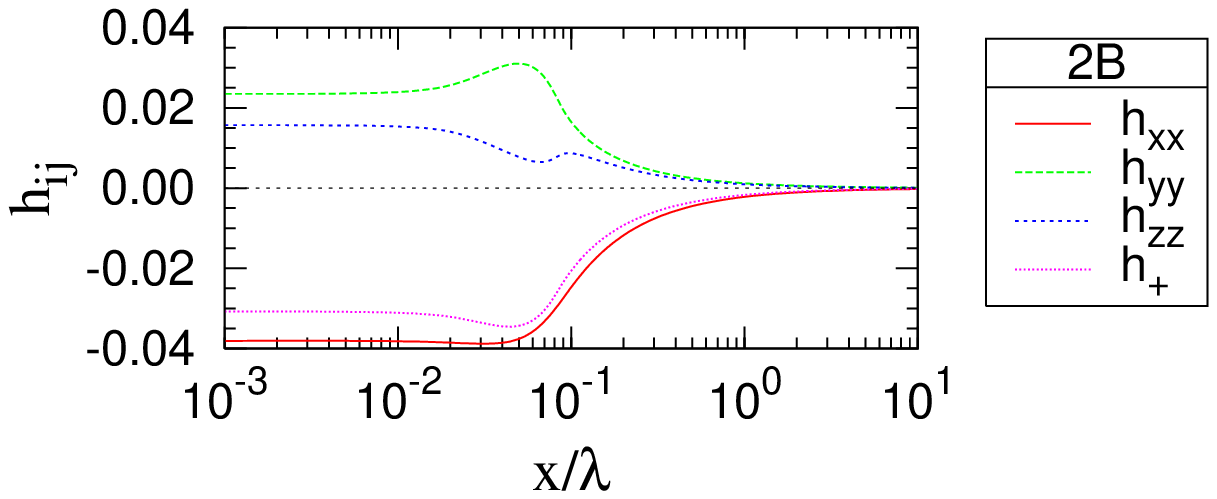}
\end{center}
\end{minipage} 
&
\begin{minipage}{.5\hsize}
\begin{center}
\includegraphics[height=35mm]{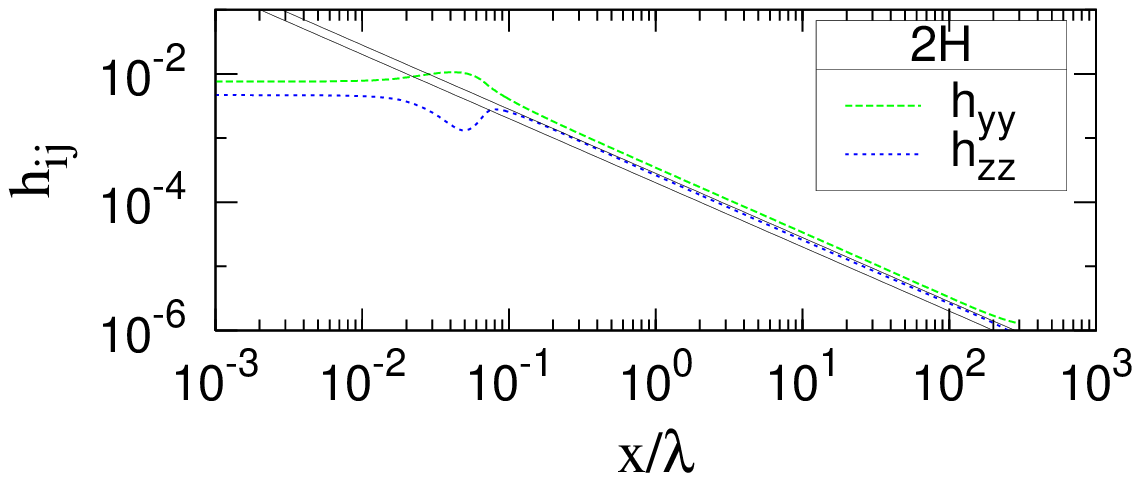}
\includegraphics[height=35mm]{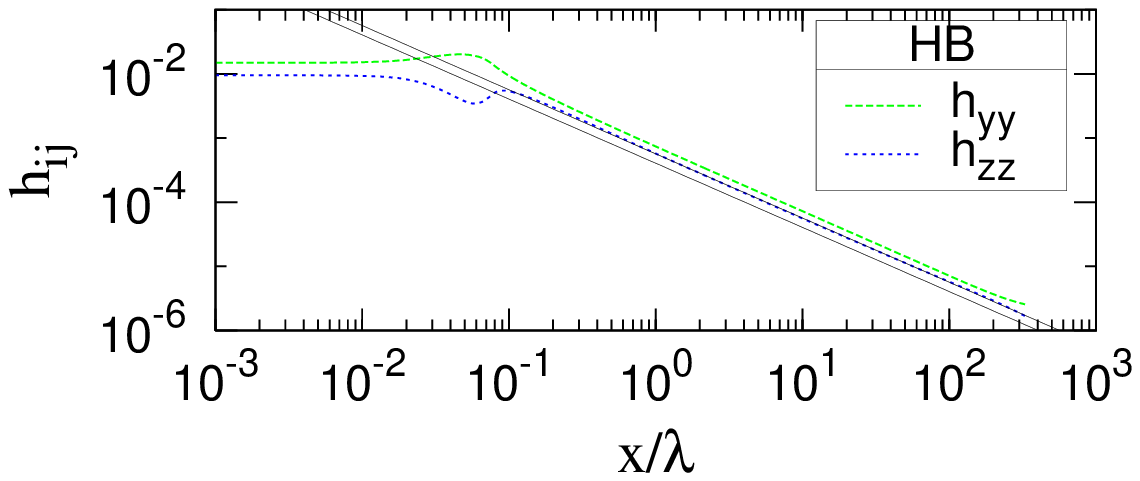}
\includegraphics[height=35mm]{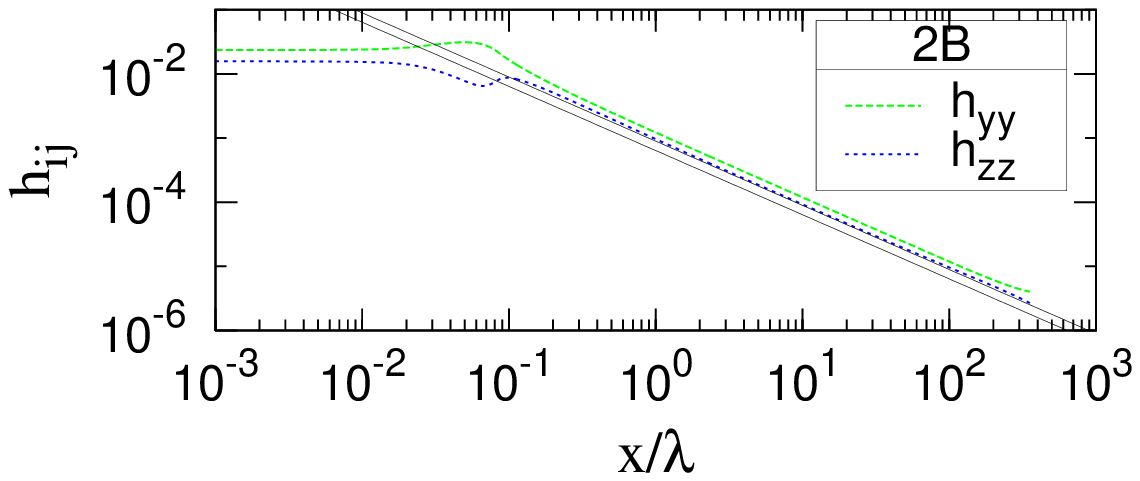}
\end{center}
\end{minipage} 
\end{tabular} 
\caption{
Selected components of $h_{ij}$ along the $x$-axis 
of the WL solutions with the orbital radius $d/R_0 = 1.5$ 
for parametrized EOS 2H, HB, and 2B, from top to bottom panels of 
both sides respectively.  Left (right) panels 
are Log-Linear (Log-Log) plots, where the $x-$axis 
is normalized by $\lambda :=\pi/\Omega$.  Upper and lower thin 
solid lines in the right panels are, respectively, 
the $h_{yy}$ and $h_{zz}$ components of the 
asymptotic solutions (\ref{eq:2PN_hij_asymp}) 
of two point mass, $M_1 = 1.35M_\odot$ each, separated
as the numerical solutions, $d/R_0 = 1.5$.  
$h_+$ in left panels is defined by $h_+:=(h_{xx}-h_{yy})/2$.
}
\label{fig:hijxaxis}
\end{figure*}
%
%
%
\subsection{Behavior of $h_{ij}$ for selected solutions}

Quasi-equilibrium solutions of irrotational 
BNS are calculated for the
various sets of EOS parameters summarized in Table \ref{tab:EOS}.  
As an example of the WL solutions, 
we present in Fig.\ref{fig:hab_contour} contours of selected 
components of $h_{ij}$ 
for the parametrized EOS HB, 
with orbital radius $d/R_0 = 1.5$, where $R_0$ is the coordinate 
radius (half the diameter) of the neutron star along 
the $x$-axis.  
For a qualitative comparison, contours are also shown for 
the leading order terms 
$\Od(r^{-1})$ of the asymptotic solution of $h_{ij}$ 
in a second order post-Newtonian (2PN) approximation 
with maximal slicing and a transverse-traceless gauge for 
$h_{ij}$, as derived in \cite{ASF96} (see, Eq.~(5.30)),
namely 
\beqn
h_{ij} &=& \frac1{r}\left\{
\frac14 I_{ij}
+\frac34 n^k (n^i I_{kj}+n^j I_{ki})
-\frac58 n^i n^j I_{kk}
\right.
\nonumber\\
&&\left.
+ \frac38 n^i n^j n^k n^l I_{kl}
+ \frac18\dl_{ij}I_{kk}
- \frac58\dl_{ij}n^k n^l I_{kl}
\right\}
\nonumber\\
&&
+ \Od(r^{-2})
\label{eq:2PN_hij_asymp}
\eeqn
where
\beq
I_{ij} =  \int \rho x^i x^j d^3x, \ \ \mbox{and}\ \ 
n^i = \frac{x^i}{r}. 
\eeq
In the quadrupole integrals $I_{ij}$, 
we substituted two $1.35 M_\odot$ point masses, 
separated by the same coordinate length as 
the above WL solution.   
The region shown in these figures does not extend 
far enough to have asymptotic behavior, though the 
contours qualitatively agree.

In Fig.~\ref{fig:hijxaxis}, selected components of $h_{ij}$
are plotted along the $x$-axis for the cases with 
parametrized EOS 2H, HB, and 2B, from top to the bottom panels.
In each case the the orbital radius is again $d/R_0 = 1.5$.  
In our models, the gravitational mass 
of the corresponding spherical star is $M_1 = 1.35M_\odot$ 
and $\compa$ of each EOS 
increases in the order of 2H, HB, 2B (see Table \ref{tab:EOS}), 
which is reflected by the increasing amplitude of $h_{ij}$.  
Here and after, the {\em compactness} of each component star in 
the binary system means the value of $\compa$ for a single spherical 
star with the same rest mass.  

In the right panels, corresponding to the left panels, 
log-log plots of the $h_{yy}$ and $h_{zz}$ components 
are shown up to the boundary of the computational domain.  
Upper and lower thin 
black lines in the right panels are, respectively, 
the $h_{yy}$ and $h_{zz}$ components of the 
asymptotic solutions (\ref{eq:2PN_hij_asymp}) 
of two point masses.  
These lines do not exactly match the $\hijd$
countours of the corresponding  
of numerical solutions for several reasons, 
including finite-size and higher order post-Newtonian effects. 
However, the lines shift systematicaly from the numerical $\hijd$, 
which suggest that the numerical $\hijd$
scales properly in the asymptotic region (as well as in the 
near zone) as the compactness increases.

\subsection{Quasi-equilibrium sequences with different compactness}
\label{Sec:2HHB2B}

A constant-rest-mass sequence of quasi-equilibrium solutions 
for irrotational BNS is considered as a model for 
the last several orbits of inspiral before merger.  
Such sequences are computed for the models with different 
EOS parameters listed in Table \ref{tab:EOS}.  The fixed rest mass 
of each model is that of a spherical star whose gravitational 
mass is $M_1 = 1.35M_\odot$.  Quantities of the spherical star 
for each model are also presented in the same Table.  

\begin{figure}
\begin{center}
\includegraphics[height=60mm]{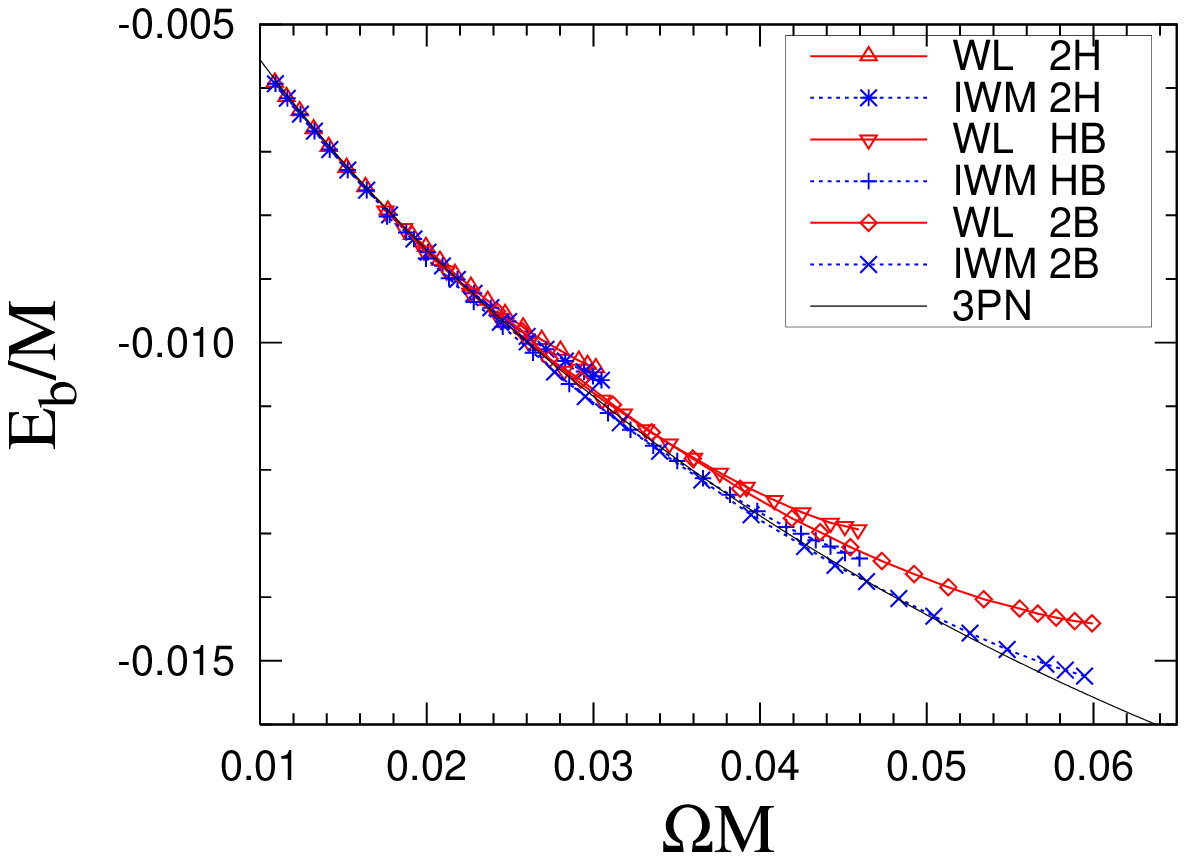}
\includegraphics[height=60mm]{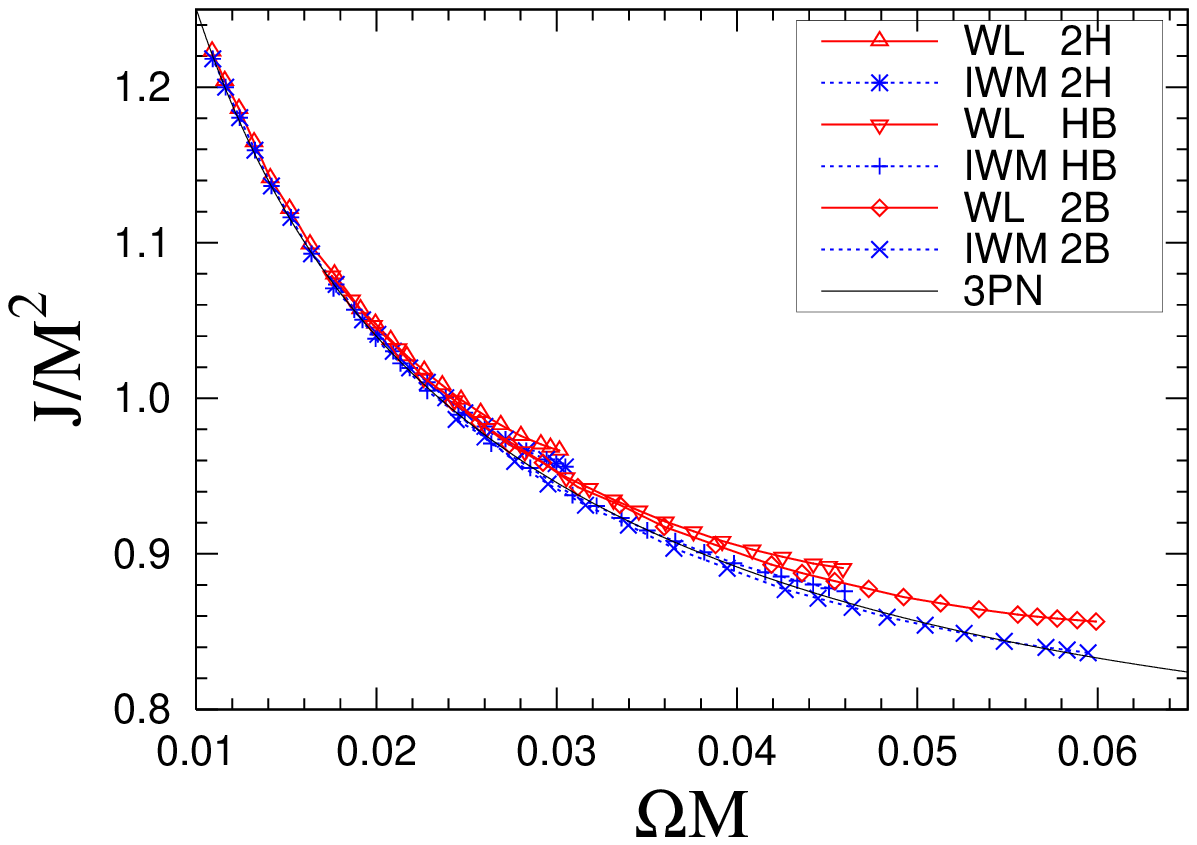}
\includegraphics[height=60mm]{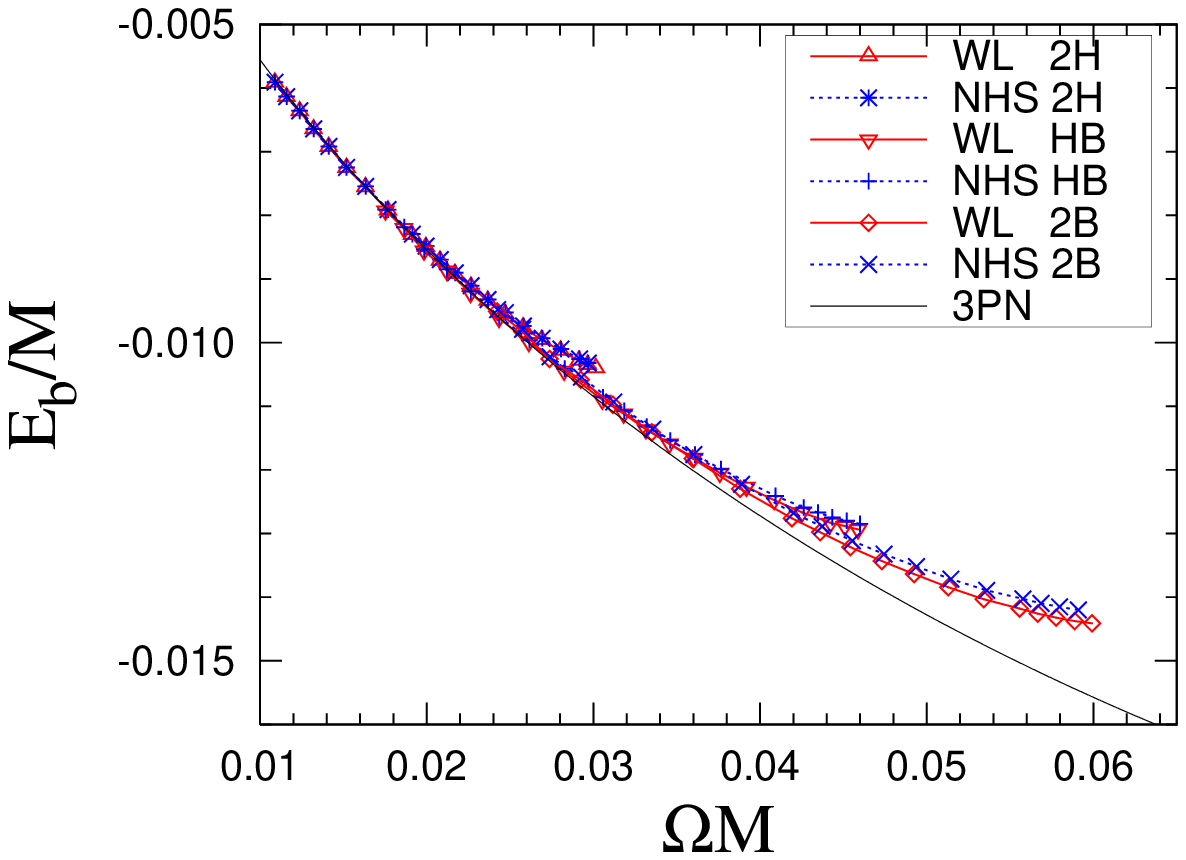}
\caption{Plots of the WL, NHS, and IWM sequences for 
the parametrized EOS 2H, HB, and 2B.  
Top panel: Binding energy 
$E_{\rm b} = \Madm - M$ normalized by $M = 2M_1$ 
with respect to the normalized angular velocity 
$\Omega M$ of the WL and IWM sequences.  
Middle panel: Total angular momentum $J$ 
normalized by $M^2$ of the WL and IWM sequences. 
Bottom panel: Normalized binding energy 
of the WL and NHS sequences.  
In each panel, a thin solid curve corresponds to 
that of the 3PN approximation.}
\label{fig:seq_2HHB2B}
\end{center}
\end{figure}
\begin{figure}
\begin{center}
\includegraphics[height=55mm]{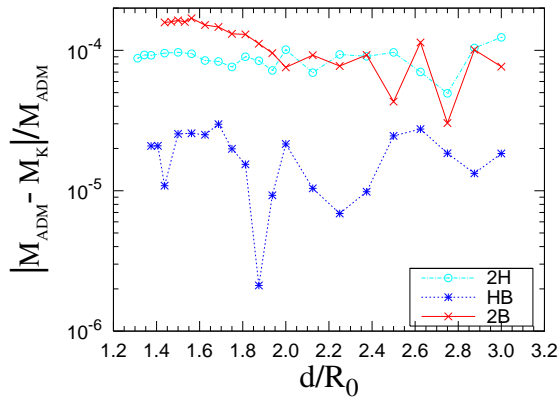}
\caption{Fractional differences of $\Madm$ and $\MK$ 
with respect to the orbital radius $d/R_0$ 
of the WL sequences for parametrized EOS 2H, HB and 2B.  
}
\label{fig:virial}
\end{center}
\end{figure}
%
%
%
\begin{figure}
\begin{center}
\includegraphics[height=60mm]{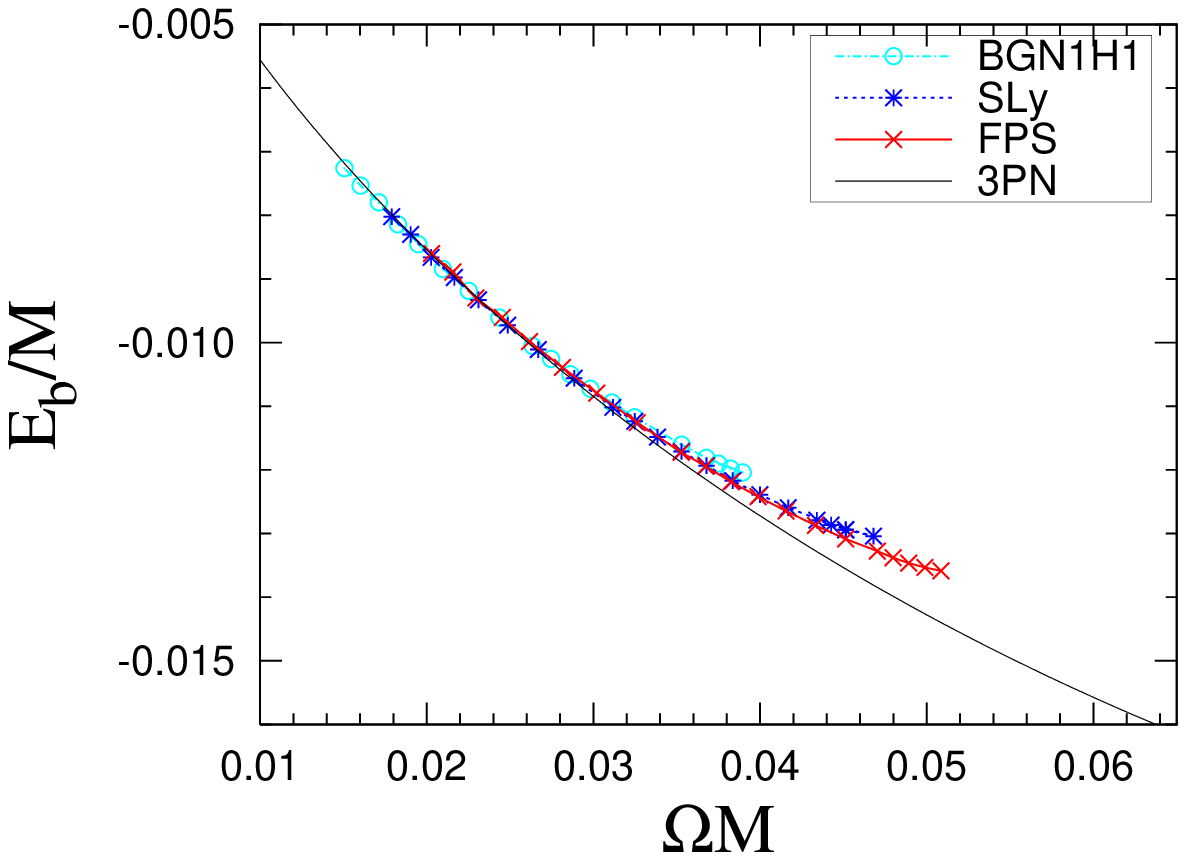}
\includegraphics[height=60mm]{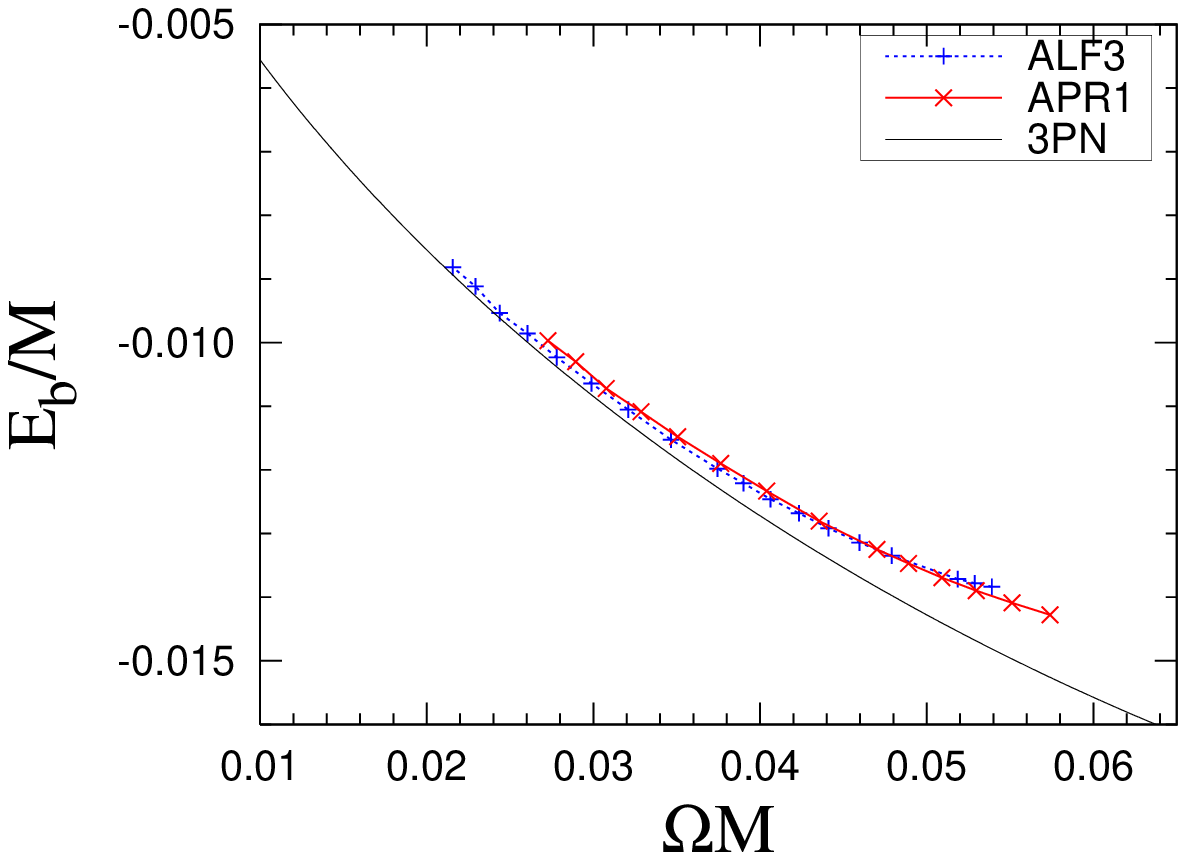}
\caption{
Binding energy $E_{\rm b}$ of WL sequences for 
the four parameter fitted EOS.  
Top panel: for the EOS, BGN1H1, SLy, and FPS.  
Bottom panel: for the EOS, ALF3, and APR1.  
}
\label{fig:seq_4pEOS}
\end{center}
\end{figure}
%
%

In Fig.\ref{fig:seq_2HHB2B}, 
the binding energy $E_{\rm b}:=\Madm - M$ and 
the total angular momentum $J$, normalized by 
twice the gravitational mass of the spherical star 
$M = 2M_1$, are plotted for models 2H, HB, and 2B.  
In the top and middle 
panels, the results of the WL sequences are compared 
with the results of IWM sequences and of non-spinning 
point particles in 3PN circular orbits.  
Clearly, the IWM sequences coincide with the 3PN curve up to 
smaller separation (larger $\Omega M$), 
whereas the WL sequences 
significantly deviate from the 3PN sequence.  
As the compactness (in this case from 2H to 2B) increases, the curves of 
the IWM sequence around the smallest separation 
come closer to the 3PN curve.  In contrast to this, 
deviations of the WL sequences from the 3PN curve are even 
larger for the larger compactness.

In the bottom panel of the Fig.\ref{fig:seq_2HHB2B}, 
the binding energy $E_{\rm b}:=\Madm - M$ 
of the WL sequences are compared with the results of 
the NHS sequences.  Clearly, the difference in 
the binding energy of two formulations is less than a percent; 
that is, the WL solutions almost coincide with the 
helically symmetric solution in the near zone.

In \cite{SUF04}, we have derived asymptotic conditions for 
equality $\Madm=\MK$ of 
the ADM and Komar masses \cite{Komar5962}, 
which is related to the relativistic virial 
relation for the equilibrium 
\cite{REMARK},   
\beq
\int x^a \gamma_a\!{}^\alpha \nabla_\beta\Tba \sqrt{-g}d^3x =0.  
\label{eq:virial_eq}
\eeq
In the WL/NHS formulation, the asymptotic fall-off of 
each field is sufficiently fast to enforce the equality.  
In Fig.~\ref{fig:virial}, we evaluate the values of 
the fractional differences 
$|\Madm - \MK|/\Madm$ for the WL sequences 
with the parametrized EOS 2H, HB, and 2B.  
The plots show that the differences are less than 
$2\times 10^{-4}$.  The compactness increases in 
the order of 2H, HB, and 2B; the fractional 
differences, however, do not necessarily increase with increasing 
compactness in this range $\compa\alt 0.2$.  
The fact that the fractional difference is well controlled for 
these sequences is evidence that the binding energy in 
Fig.\ref{fig:seq_2HHB2B} is calculated accurately.  
The virial relation Eq.~(\ref{eq:virial_eq}), 
normalized by $\Madm$, is also calculated to examine the accuracy of 
the numerical solutions, whose absolute value is about $0.5 \sim 1$
times that of the fractional difference of two masses.

\subsection{Quasi-equilibrium sequences with 
four-parameter fitted EOS}
\label{Sec:4pEOS}

In the paper \cite{Read:2008iy}, optimal values for the parameters 
of four-parameter fitted EOS have been derived for 
34 candidates of the neutron-star EOS (17 selected EOS of 
nuclear matter with varied parameters).  
We choose five representative EOS, which are 
SLy \cite{SLy}, APR1 \cite{APR}, FPS \cite{FPS}, 
BGN1H1 \cite{BGN1H1}, and ALF3 \cite{ALF}.  
The first three are made only from normal nuclear matter, 
while BGN1H1 involves a mixed phase with hyperons, 
and ALF3 with quarks.  For the latter two EOS, the value of 
$\Gamma$ becomes smaller in the mixed phase with the exotic matter 
at a few times above nuclear density \cite{Read:2008iy}.  
However, BGN1H1 is a stiff EOS having the largest $p_1$ 
among them, and hence the core of the mixed phase is not 
large for the mass $M_1 = 1.35 M_\odot$.  

In Fig.\ref{fig:seq_4pEOS}, the binding energy $E_{\rm b}$ of 
the WL sequences for these parametrized EOS are plotted.  
As in the case of the one-parameter 
parametrized EOS in Sec.\ref{Sec:2HHB2B}, 
the sequences with higher compactness $\compa$ 
extend to higher values of $\Omega M$.  
Also, the WL sequences deviate from the 3PN curve at 
larger $\Omega M$.  Among these EOS, APR1 is the softest, 
giving the most compact neutron-star model; and the corresponding binary 
sequence reaches the highest value, $\sim 0.058$, of $\Omega M$.  
However, as seen in the bottom panel of Fig.~\ref{fig:seq_4pEOS}, 
the binding energy curve of APR1 is slightly off from the 3PN 
curve even for the smaller $\Omega M$ of the sequence.  
In our neutron-star code, using a finite difference scheme, 
the core of the neutron star is covered by fewer grid points 
in the central coordinates when the binary separation becomes 
larger and the neutron stars more compact; 
this may increase the numerical errors.  
We plan to incorporate a binary computation in the new code 
\cite{Huang:2008vp}, in which enough grids are maintained, 
to densely to cover the neutron star, irrespective of the binary separation or 
neutron-star radius.  The results of the APR1 curve as well as 
more compact binary sequences will be studied using the new code.  

In \cite{Read:2009yp}, the gravitational waveform
computed from inspiral simulations has been analyzed 
to estimate the accuracy with which gravitational wave 
observations can constrain neutron-star radius, an EOS 
parameter correlated with the departure from point-particle 
inspiral.  A promising result is 
that the neutron-star radius can be constrained 
to $\dl R \sim 1 \mbox{km}$ 
for an interferometric detector with the sensitivity of Advanced LIGO, in 
either a broadband configuration or a narrowband with peak 
sensitivity around 1150Hz.  
This suggests that the successful observations of gravitational 
waves may exclude even a couple of EOS shown in 
Fig.\ref{fig:seq_4pEOS}.

\subsection{Comparison of the orbital phase in the last several orbits}
\label{sec:orbits}

In this section, we approximately determine the orbital evolution in
the late inspiral phase up to the onset of merger using the
quasi-equilibrium sequences computed in the previous section.  
To construct a quasi-equilibrium sequence, one assumes that each BNS evolves adiabatically along the sequence, that the radial velocity is much smaller
than the orbital velocity.  Given the rest mass and the EOS,
each quasi-equilibrium sequence is defined by one parameter: 
The total energy and angular momentum of the binary system along 
a sequence are parametrized by the orbital angular velocity as 
$E(\Omega)$ and $J(\Omega)$. 

The time evolution of the angular velocity then becomes 
\beqn
\frac{d\Omega}{dt}=\biggl(\frac{dE}{d\Omega}\biggr)^{-1}
\frac{dE}{dt} \equiv F(\Omega)^{-1}.  
\label{dtdomega}
\eeqn
For the gravitational wave luminosity, $dE/dt$, 
we adopt the 3.5PN formula for two point 
masses \cite{PN}. Tidal deformation of the neutron stars 
in close orbits makes the attractive force between two stars 
stronger, and hence it accelerates the orbital velocity, 
resulting in the enhancement of the gravitational wave luminosity. 
Thus, the 3.5PN formula for the luminosity is likely to 
underestimate that of the BNS. However, this effect 
plays an important role only for the last $\sim 1$ orbit, and 
for most of the late inspiral orbits, the 3.5PN formula is 
a good approximation.

Numerical integration of Eq. (\ref{dtdomega}) provides the relation between 
$t$ and $\Omega$ from 
\beqn
t = \int d\Omega F(\Omega). \label{dtdw}
\eeqn
From this, the angular velocity as a function of time, $\Omega(t)$, is 
obtained. Using this relation, 
we can also compute the approximate orbital phase evolution by
\beqn
N=\frac{1}{2\pi} \int \Omega(t) dt. 
\eeqn
We note that the numerical model with the maximum value of $\Omega$ 
for each sequence presented in this paper does not exactly, 
but does approximately, correspond to a solution at the closest orbit.  
We stop the integration of Eq. (\ref{dtdw}) when $\Omega$ reaches its
maximum. 

\begin{figure}[t]
\begin{center}
\includegraphics[height=60mm]{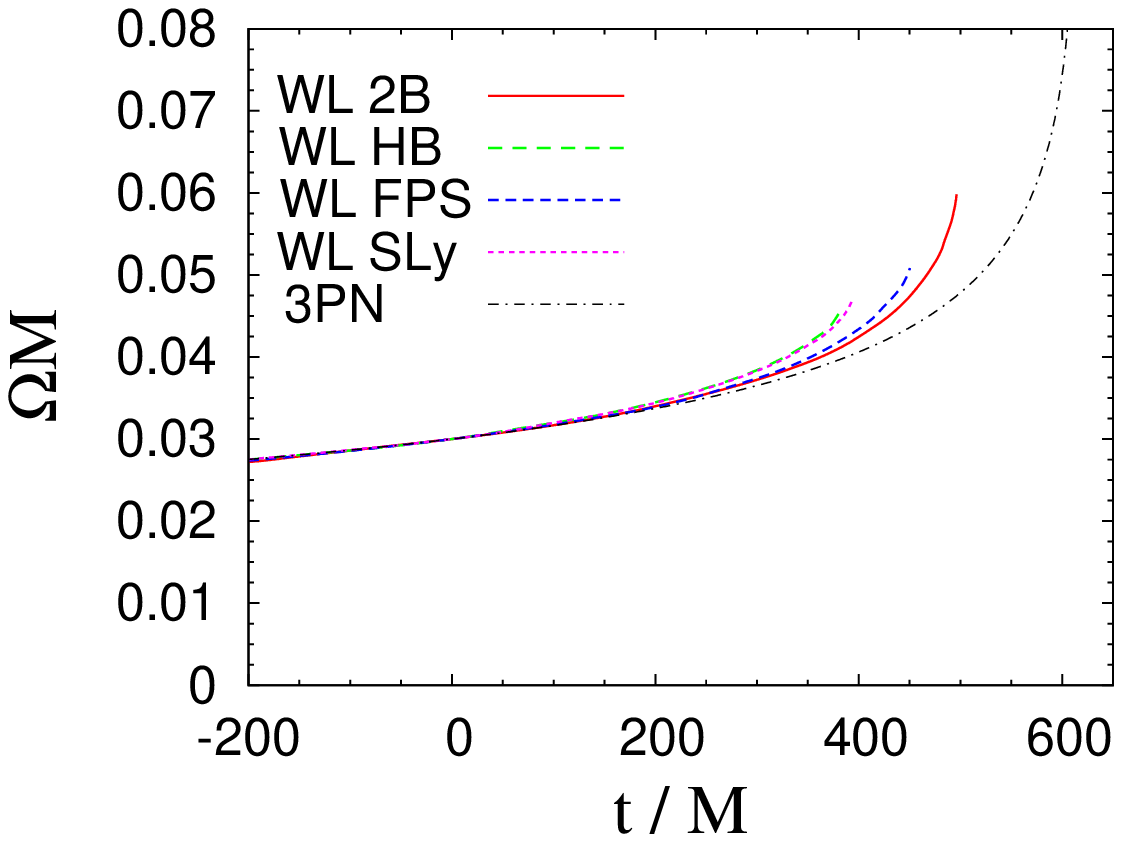}
\includegraphics[height=60mm]{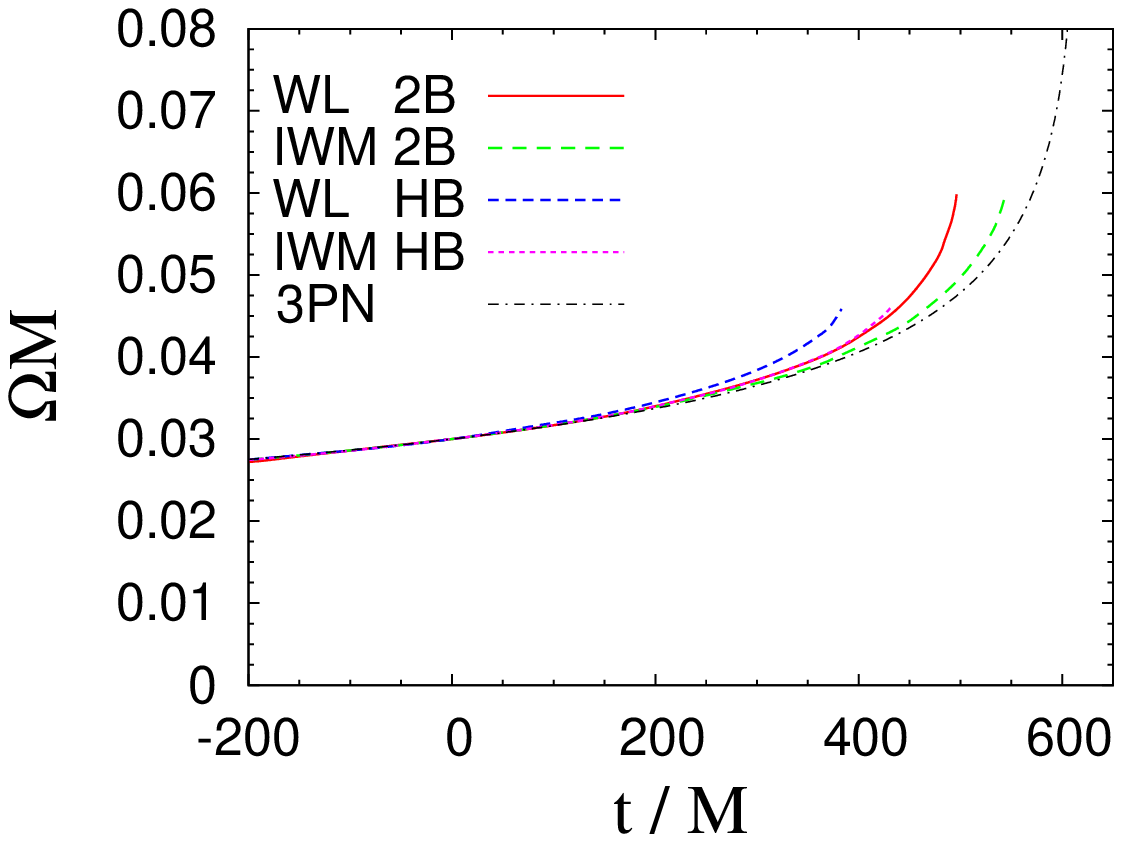}
\caption{
Top panel: Orbital angular velocity, $\Omega$, as a function of time
for EOS 2B, HB, FPS, and SLy of the WL sequences.  
Bottom panel: 
Same as the top panel 
but for EOS 2B and HB of the WL and IWM sequences.  
For the both panels, 
the results by the Taylor-T4 formula are also plotted.  
The units of $\Omega$ and time are $M^{-1}$ and $M$,
respectively.  For comparing the results, the time axis 
is shifted such that $\Omega M=0.03$ is aligned at $t=0$ 
for all the curves.
\label{ooo1}}
\end{center}
\end{figure}

In the top panel of Fig.~\ref{ooo1}, $\Omega M$ is plotted as 
a function of time for EOS 2B, HB, FPS, and SLy in the WL formulation.  
In the bottom panel of Fig.~\ref{ooo1}, the
results for 2B and HB, calculated in both the WL and IWM formulation, 
are compared.  We also plot the results of two point masses, derived 
from the Taylor-T4 formula \cite{TT4}.

The top panel of Fig.~\ref{ooo1} shows that for the small values of $\Omega$, all
the curves approximately agree, irrespective of the EOS.  This is
natural because for such small values, tidal deformation does
not play an important role and orbital velocity is sufficiently
small ($v < 0.3c$) that the
post-Newtonian formula (Taylor-T4 formula) with the point-particle
approximation should be an excellent approximation.

By contrast, the values of $\Omega(t)$ computed from the numerical
sequences deviate from those given by the Taylor-T4 formula for $\Omega M \agt
0.035$--0.04, for all of the EOS and all formulations used to 
compute the quasiequilibria.  This is due to the tidal 
deformation of the neutron stars; the rate of change of 
the energy as a function of $\Omega$ approaches zero for 
the close orbits, as seen in Figs.~\ref{fig:seq_2HHB2B} and \ref{fig:seq_4pEOS}. 
This deviation occurs at more distant orbits for
less compact neutron stars (i.e., for the stiffer EOS), indicating,
as expected, that one can extract from the curve $\Omega(t)$ 
a characteristic of the component neutron stars related to their 
compactness and a corresponding parameter of the EOS. 

The bottom panel of Fig.~\ref{ooo1} shows that the curves $\Omega(t)$ computed by 
the WL and IWM formulations are significantly different, as expected from the
results of $E(\Omega)$.  In the case that the IWM formulation is adopted,
the merger time is overestimated by $\sim 50M$, which is a quite a large
factor. This suggests that the results in the IWM formulation do not work 
well for predicting the evolution of the last several orbits before the
onset of merger.

\begin{figure}[t]
\begin{center}
\includegraphics[height=60mm]{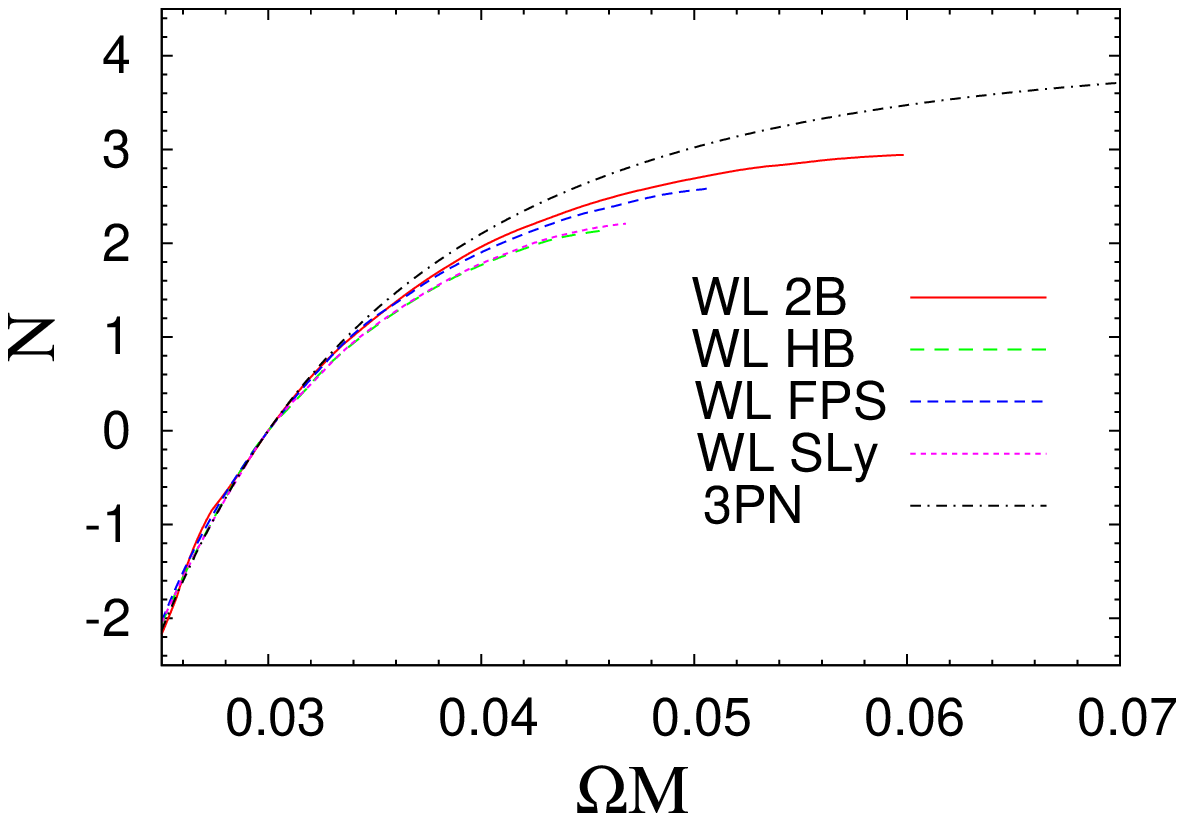}
\includegraphics[height=60mm]{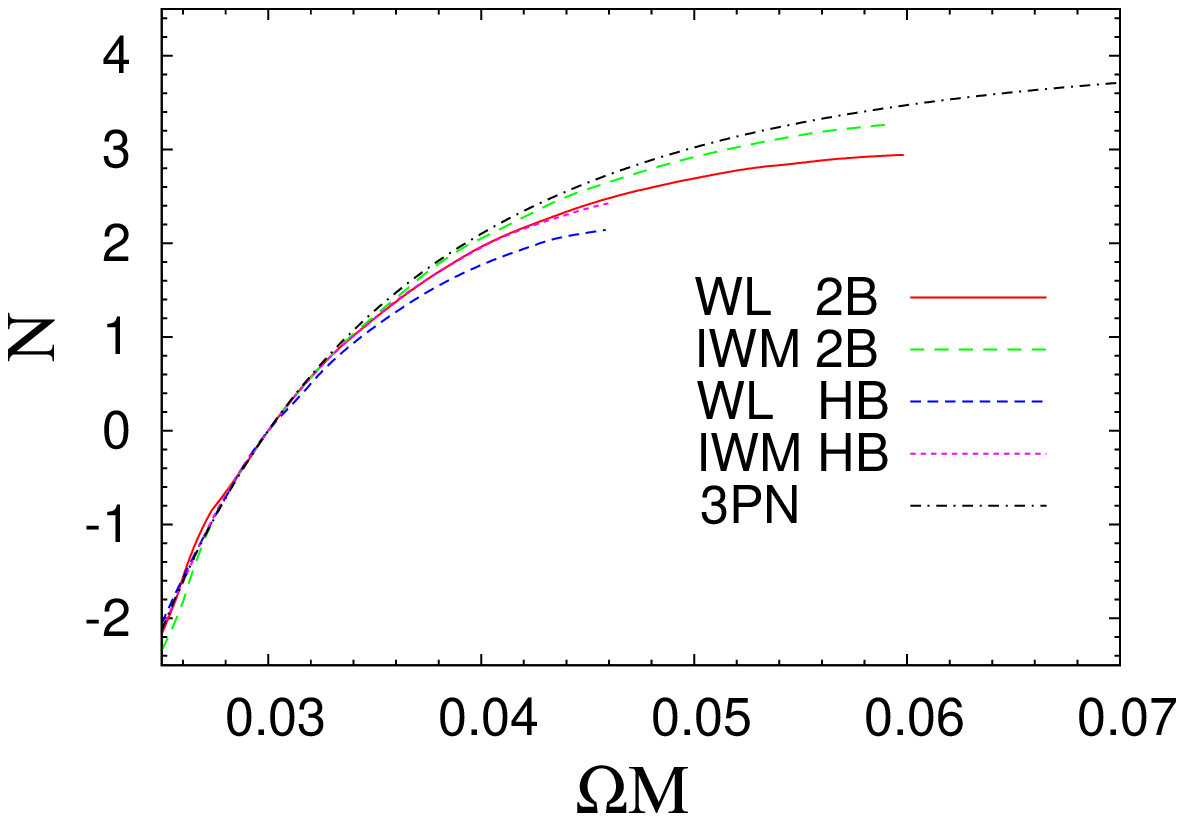}
\caption{
Top panel: Orbital cycle, $N$, as a function of time for EOS 2B,
HB, FPS, and SLy of the WL sequences.  
Bottom panel: The same as the above but for EOS
2B and HB of the WL and IWM sequences.  
For the both panels, 
the results of the Taylor-T4 formula are also plotted. 
For comparing the results, the time axis is shifted such that 
$N=0$ is aligned at $\Omega M=0.03$ for all the curves.
\label{ooo2}}
\end{center}
\end{figure}

In Fig.~\ref{ooo2}, we plot the curves of $N$ as a function of
$\Omega M$; the top panel is for EOS 2B, HB, FPS, and SLy in the
WL formulation and the bottom for EOS 2B and HB in the WL and 
IWM formulation.
The top panel shows that the number of orbital cycles in the
late inspiral phase depends strongly on the EOS. For a soft EOS,
e.g., EOS 2B, in which the compactness of the neutron star is largest,
the number of cycle is largest. 
By contrast, for a stiff EOS such as SLy, the number 
of cycles may be smaller by $\sim 1$ than that for EOS 2B.

In bottom panel of Fig.~\ref{ooo2}, the results for the 
number of cycle calculated from different formulations 
are compared.  As expected from 
the results for $\Omega(t)$, the IWM formulation overestimates 
the number of cycles. The error $\Delta N$ is $\sim 0.5$ for 
the EOS 2B; i.e., one cycle of gravitational waves would be 
overestimated.

\section{Discussion}

The deviations of the binding energy and total angular momentum
of WL/NHS sequences from
the 3PN point-particle sequence as well as from the IWM sequences
are likely to be due to the tidal deformation of neutron stars 
in the binary system coupled with general relativistic effects.
As the compactness of the component neutron stars increases,
the deviation from the 3PN sequence at a certain value
of $\Omega M$ decreases -- WL/NHS sequences become closer
to the point-particle sequence, but not by as much as
the IWM sequences do.
It has been believed that, as the compactness of the component 
neutron stars  
increases, the behavior of the binding energy and angular momentum 
of binary sequences more closely approximates that of point masses.  
This is found in the results of IWM sequences but to a lesser extent 
in the WL/NHS sequences.  
The behavior of the IWM sequence was interpreted 
as the effacing of the tidal effects due to the 
strong gravity: that is, 
as the compactness increases, the sequences of binary 
neutron-star solutions become 
much closer to the sequences of two point masses, because 
the tidal effect is masked by the stronger self-gravity 
of each component star.  
However, the results of WL/NHS sequences suggest that such 
effacing of the tidal effect seen in IWM sequence 
is an artifact of the conformally flat 
approximation, at least for the case of equal mass binary 
neutron stars.

In the WL/NHS formulations, all components of Einstein's equation 
are solved without approximation on a initial hypersurface, while in 
the IWM formulation, some terms of second post-Newtonian order are 
truncated.  As discussed in \cite{Uryu:2005vv} 
the difference between the IWM and WL/NHS formulations in the 
binding energy $E_b$ is estimated at second post-Newtonian order 
as $M h_{ab} v^a v^b$, where the magnitude of the orbital velocity 
$v^a$ is typically $v\sim 0.34(\Omega M/0.04)^{1/3}$.  
Since $h_{ab}$ is $\Od (v^4)$, the order of the difference in 
the binding energy is given by $\Delta E_b/M = \Od (v^6)\sim 10^{-3}$, 
and a larger deviation as $v$ becomes larger for more compact 
sequences is expected.  This estimate is consistent with our results 
shown in Fig.~\ref{fig:seq_2HHB2B} and \ref{fig:seq_4pEOS}.  
Note also that the tidal effect is larger 
for the EOS with a larger $\Gamma$ as we used in our computations.  
So far, our WL/NHS codes have passed several code tests (as 
have the IWM codes), and results of two independent WL codes agreed for a 
BNS sequence with $M_1/R = 0.17$ as shown in \cite{Uryu:2005vv}.
These results support our argument that the WL/NHS results 
accurately correct the IWM results.  
A computation of quasi-equilibrium BNS sequences using 
a totally different numerical method, such as the fully constrained 
scheme \cite{BGGN04}, would be a helpful additional check.  

We think our results suggest that the circularity of orbit is more 
accurately enforced on a WL/NHS sequence than a IWM sequence.  
However, in such quasi-equilibrium sequences, some important features 
of realistic inspirals are ignored.  Those include the radial 
velocity due to gravitational radiation reaction at 2.5PN order 
that is likely to be comparable to the 2PN terms during the 
last few orbits where the neutron-star velocity is of order 
$v\sim 0.1$, and a tidal lag angle of about $10 - 20$ degrees 
that is found in inspiral simulations.  Therefore, a caveat is 
that estimates of the merger time and orbital cycles 
using quasi-equilibrium sequences shown in Sec.~\ref{sec:orbits} 
involve errors due to ignoring these effects.  

Recently, several groups have developed methods to 
treat the general relativistic tidal deformations 
analytically \cite{GRLove}.  
Comparison of these analytic results and 
the present results for WL/NHS sequences 
may be useful in calibrating  
the binding energy or the total angular momentum 
of the quasi-equilibrium sequence in the regime where 
the relativistic tidal effects become important.  
Finally, by combining the analytic and numerical results, 
more accurate quasi-equilibrium models for 
the late inspirals may be constructed \cite{jocelyn}.

The WL/NHS formulations can be also used to 
construct models of rotating neutron stars.  
In \cite{Lin:2006sm}, axisymmetric rotating 
relativistic stars are computed using the fully constrained 
formulation with maximal slicing and the generalized 
Dirac gauge conditions \cite{BGGN04}.  
Those solutions agreed with the ones calculated 
using a stationary axisymmetric metric with 
the additional discrete symmetry of the simultaneous transformation, 
$t\rightarrow -t$ and $\phi\rightarrow -\phi$.  
The WL/NHS formulations include more general stationary 
axisymmetric spacetimes, which do not depend on the additional symmetry. 
Therefore, the WL/NHS formulations can be applied, for example, 
to rotating neutron stars that may have both toroidal 
and poloidal components of the magnetic fields 
as well as meridional circulation.  Even in this case, 
the WL/NHS formulation 
can be used to compute exact equilibria that are more 
general than those calculated in \cite{KY08}.  We plan to 
extend our codes to compute relativistic rotating stars and 
binary systems that each include strong magnetic fields.

\acknowledgments
This work was supported by 
JSPS Grant-in-Aid for Scientific Research(C) 20540275 
and (B) 21340051, 
MEXT Grant-in-Aid for Scientific Research
on Innovative Area 20105004, 
NSF grants Nos.~PHY0071044 and PHY0503366, 
NASA grant No.~NNG05GB99G, 
and ANR grant 06-2-134423 \emph{M\'ethodes 
math\'ematiques pour la relativit\'e g\'en\'erale}.  
KU and EG acknowledge a support from JSPS Invitation Fellowship 
for Research in Japan (Short-term) and the invitation program 
of foreign researchers at Paris observatory.

\appendix
\section{Basic equations}
\label{secA:eqs}

In this Appendix, the system of equations 
used in WL/NHS codes is presented in detail.  The equations 
include all components of Einstein's equation, the first integral of 
the relativistic Euler equation, and the rest mass conservation 
equation for the irrotational flow.  
The WL/NHS formulations are based on 
\cite{SUF04,BGGN04,YBRUF06}.

\subsection{Conventions}

As mentioned in Sec.\ref{sec:3plus1}, the 3+1 decomposition is 
applied to the spacetime $\cal M$ in the WL/NHS formulations.  
First, several definitions for the quantities relating 
to the spatial geometry are introduced.  

\subsubsection{Connections}

The spatial metric $\gmabd$, a conformally rescaled 
spatial metric $\tgmabd$, and a flat metric 
$f_{ab}$ are associated with the derivatives 
$D_a$, $\tD_a$, and $\zD_a$, respectively.  We introduce 
the conformal rescaling by $\gmabd = \psi^4 \tgmabd$, 
whose determinant $\tgamma$ is equal to that of the flat 
metric $f$, $\tgamma = f$, to specify the decomposition 
of the spatial metric uniquely. 
Covariant derivatives $D_a$ and $\tD_a$ are related by 
\beq
  D_b X^a = \tD_b X^a+ \tC^a_{bc}X^c, 
\eeq
where $X^a$ is a spatial vector, 
and a coefficient $\tC^c_{ab}$ is written 
\beqn
\tC^c_{ab} 
&=& \frac12\gamma^{cd}
(\tD_a\gamma_{db} +\tD_b\gamma_{ad} -\tD_d\gamma_{ab} ) 
\nonumber\\
&=& \frac{2}{\psi}\,(\tgamma^c\!{}_b \tD_a \psi + \tgamma^c{}_a \tD_b \psi
-\tgamma_{ab} \tgamma^{cd}\tD_d \psi).  
\eeqn
Also, $\tD_a$ and $\zD_a$ are related by 
\beq
\tD_b X^a = \zD_b X^a+  C^a_{bc}X^c,
\eeq
where $C^c_{ab}$ is written 
\beqn
C^c_{ab} 
&=& \frac12\tgamma^{cd}
(\zD_a\tgamma_{db} +\zD_b\tgamma_{ad} -\zD_d\tgamma_{ab} )
\nonumber\\
&=& \frac12\tgamma^{cd}
(\zD_a h_{db} +\zD_b h_{ad} -\zD_d h_{ab} ).
\eeqn

A trace of $C^c_{ab}$
\beq
C^b_{ba} 
\,=\, \frac12 \tgamma^{bc} \zD_a \tgamma_{bc}
\,=\, \frac1{\sqrt{\tgamma}}\zD_a \sqrt{\tgamma}, 
\eeq
and the condition $\tgamma=f$ that specifies the conformal decomposition 
imply $C^c_{ca} =0$ and hence 
$\tD_a \psi = \zD_a \psi$.  
The relations 
\beq
\tgmabu C^c_{ab} \,+\, \tgamma^{bc} C^a_{ab} 
\,+\, \zD_a \tgamma^{ac}
\,=\,0 
\eeq
and $\tgamma=f$, and the Dirac gauge condition $\zD_b\tgmabu=0$ 
imply $\tgmabu C^c_{ab}=0$.

\subsubsection{Conformally rescaled extrinsic curvatures}
\label{sec:Kab}

The form of the extrinsic curvature $\Kabd$ is discussed in 
Sec.\ref{sec:WL_HS}.  In the equations for our 
numerical code, it is decomposed in terms of the trace 
$K = \gmabu\Kabd$ and the tracefree part $\Aabd$,
\beq
\Kabd = \Aabd + \frac13 \gmabd K.
\eeq
The conformally rescaled tracefree part $\tAba$ is defined as 
\beq
\tAba=\Aba,
\eeq 
and its index is lowered (raised) by 
$\tgmabd$ ($\tgmabu$).

We define $L_X \gmabd$ as 
the tracefree part of $\Lie_X \gmabd$, 
where $X^a$ is a spatial vector on $\Sigma_t$,  
\beqn
L_X \gmabd
&=& \Lie_X\gmabd - \frac13 \gmabd \gmcdu\Lie_X \gmcdd
\\
&=&
D_a X_b+D_b X_a -\frac23\gamma_{ab} D_c X^c
\eeqn
The r.h.s.~of this equation is a conformal Killing 
operator, and its conformally rescaled version is defined by  
\beq
\tilde L_X \tgmabd \,=\, \psi^4 L_X \gmabd.
\eeq
Note that a vector is rescaled, $\tilde X^a = X^a$, 
and $\tgmabd$ is used when lowering the index of the rescaled 
vector.  

When helical symmetry, $\Lie_k\gabd=0$, is imposed as in Sec.\ref{sec:WL_HS}, 
the tensors $\Aabd$ and $\tAabd$ have the forms 
\beq
\Aabd = \frac1{2\alpha} L_\omega \gmabd 
\quad \mbox{and}\quad
\tAabd = \frac1{2\alpha} \tilde L_\omega \tgmabd, 
\eeq
respectively; while for the WL formulation, 
\beq
\Aabd = \frac1{2\alpha} L_\beta \gmabd 
\quad \mbox{and}\quad
\tAabd = \frac1{2\alpha} \tilde L_\beta \tgmabd. 
\eeq
The following expression for the conformally rescaled 
$\tA_{ab}$ is used later, 
\beqn
\tAba &=& 
\frac1{2\alpha}\left(\tD_a\tbeta^b+\tD^b\tbeta_a
-\frac23\tgamma_a\!{}^b\tD_c\tbeta^c\right) 
\nonumber \\
&+& \frac1{2\alpha}\Omega\,\tgamma^{bc}\tilde{L}_\phi \tgamma_{ac} .
\label{eq:tAab}
\eeqn
The last term in the above 
\beqn
\tilde{L}_\phi\tgmabd
&=&\Lie_\phi\tgmabd - \frac13\tgmabd\tgmcdu\Lie_\phi\tgmcdd \\
&=&\tD_a\tphi_b+\tD_b\tphi_a-\frac23\tgmabd\tD_c\tphi^c,   
\eeqn
with $\tphi^a = \phi^a$ and $\tphi_a :=\tgmabd\tphi^b$, 
appears only in the helically symmetric case and is eliminated 
when the WL formulation is used.

\subsubsection{Conformally rescaled intrinsic quantities}
\label{sec:RicciRnl}

The Ricci tensor $\tR_{ab}$ of the spacelike hypersurface $\Sigma_t$ 
associated with the spatial metric $\gmabd$ is decomposed into 
terms related to the conformal factor $\psi$, $\ttR^\psi_{ab}$, 
and the conformal Ricci tensor $\ttR_{ab}$ associated with $\tgmabd$:
\beq
\tR_{ab} = \ttR^\psi_{ab} + \ttR_{ab}.  
\label{eq:riccicf}
\eeq
The first term is written 
\beqn
\ttR^\psi_{ab} 
&=& 
-\frac2{\psi}\tD_a\tD_b\psi
-\tgmabd\frac2{\psi}\tD^c\tD_c\psi
\nonumber\\
&&
+\frac6{\psi^2}\tD_a\psi\tD_b\psi
-\tgmabd\frac2{\psi^2}\tD_c\psi\tD^c\psi\, .
\label{eq:riccipsi}
\eeqn
In $\ttR_{ab}$, terms linear in $h_{ab}$ or $h^{ab}$ 
are separated as 
\beq
\ttR_{ab} \,=\,
 -\, \frac12 \zD^c\zD_c h_{ab}
 \,+\, \Rd_{ab}
 \,+\, \Rnl_{ab}, 
\label{eq:laphab}
\eeq
where $\Rd_{ab}$ includes terms linear in  
the conformal metric in the form of flat divergences  
\beqn
\Rd_{ab} &=& - \frac12(f_{ac}\zD_b F^c+f_{bc}\zD_a F^c), 
\\
F^a &:=& \zD_b \tgmabu =  \zD_b h^{ab}; 
\eeqn
non-linear terms, $\Rnl_{ab}$, are written 
\beqn
\Rnl_{ab} &=& 
-\frac12 (\zD_b h^{cd}\zD_c h_{ad}
+ \zD_a h^{cd}\zD_c h_{bd}
  + h^{cd}\zD_c \zD_d h_{ab})
\nonumber\\  
&&
  - \zD_a C^c_{cb}
  + C^c_{ab}C^d_{dc}
  - C^d_{ac}C^c_{bd}
\nonumber\\
&&-\frac12 [\,\zD_b (h_{ac}F^c)+\zD_a (h_{bc}F^c)\,] + F^c C_{c,ab}\, ,  
\label{eq:Rnonlinear}
\eeqn
where $C_{c,ab} := \tgmcdd C^d_{ab}$.  
The above expression for $\Rnl_{ab}$ can be simplified by 
applying the condition $\tgamma = f$ and the generalized Dirac 
gauge condition, implying $C^b_{ba}=0$ and $F^a=0$.  

The Ricci scalar curvature $\tR$ of $\Sigma_t$ 
is related to the conformal Ricci scalar
$\ttR := \tgmabu\, \ttR_{ab}$ by  
\beq
\tR = \frac1{\psi^4}\ttR\,-\, \frac8{\psi^5}\tD^a\tD_a \psi.  
\label{eq:scapsi}
\eeq

\subsection{Equations for the gravitational fields}

Equations used in the numerical code are 
shown below.  Although we impose the gauge conditions 
(\ref{eq:gauge_maximal}) and (\ref{eq:gauge_Dirac}),
the following equations are not restricted to these choices.
The conformal decomposition, however, is specified by 
a condition $\tgamma = f$ that is used, for example, 
to obtain the relation $\tD_a \psi = \zD_a \psi$.

\subsubsection{Hamiltonian constraint}

The projection of Einstein's equation along the normal $n^\alpha$ to 
the hypersurface yields
\beqn
&&
(\Gabd-8\pi\Tabd)n^\alpha n^\beta 
\nonumber \\
&&
\,=\, \frac12 (\tR+K^2- K_{ab} K^{ab}-16\pi\rhoH) = 0.
\eeqn

Substituting Eq.~(\ref{eq:scapsi}), we have
\beqn
&&(\Gabd-8\pi\Tabd)n^\alpha n^\beta 
=
\frac4{\psi^5}\left[
-\tD^a\tD_a\psi 
\,+\, \frac{\psi}{8}\,\ttR 
\right.
\nonumber\\
&& \left.
\,-\, \frac{\psi^5}{8}\,\left(\tA_{ab} \tA^{ab} - \frac23 K^2\right)
\,-\,2\pi\psi^5\rhoH
\right]=0.
\eeqn

The above equation is rewritten to isolate the flat Laplacian 
$\zLap \psi : = \zD^a\zD_a \psi$ on the  
l.h.s., and the other terms are treated as a source on the r.h.s., 
\beq
\zLap \psi \,=\, \SHam
\eeq
with the source $\SHam$ given by 
\beqn
\SHam &=& 
- h^{ab} \zD_a\zD_b\psi + \tgmabu C^c_{ab}\zD_c\psi
\,+\,\frac{\psi}{8}\,\ttR 
\nonumber\\
&&\,-\, \frac{\psi^5}{8}\,\left(\tA_{ab} \tA^{ab} - \frac23 K^2\right)
\,-\,2\pi\psi^5\rhoH. 
\eeqn

\subsubsection{Momentum constraint}

The momentum constraint is written in an elliptic equation 
to be solved for the covariant component of 
the conformally rescaled non-rotating shift 
$\tbeta_a:=\tgmabd\beta^b$.  We begin with 
\beqn
&&(\Gabd-8\pi\Tabd)\gmaa n^\beta 
\nonumber \\
&&\,=\, -D_b K_a{}^b+D_a K+8\pi j_a 
\nonumber \\
&&\,=\, -\frac1{\psi^6}\tD_b \left(\psi^6\tA_a\!{}^b\right)
\,+\, \frac23\tD_a K \,+\, 8\pi j_a \,=\,0, \ \ 
\label{eq:Momcon_2}
\eeqn
then substitute Eq.~(\ref{eq:tAab}) and a relation 
\beq
\tD_b\tD_a\beta^b-\tD_a\tD_b\beta^b=\ttR_{ab}\beta^b, 
\eeq
to obtain 
\beqn
&& \tD_b\tD^b\tbeta_a +\frac13\tD_a\tD_b\tbeta^b
+\ttR_{ab}\tbeta^b
+\Omega\tD_b (\tilde{L}\phi)_a{}^b
\nonumber\\
&&
+2\alpha\tA_a\!{}^b \frac{\alpha}{\psi^6}
\tD_b \left(\frac{\psi^6}{\alpha}\right)
-\frac43\,\alpha\tD_a K
-16\pi \alpha j_a \,=\,0.
\nonumber\\
\label{eq:shifteq}
\eeqn

From the first two terms of the r.h.s. of Eq.~(\ref{eq:shifteq}), 
the flat terms  
$\dis \zLap \tbeta_a +\frac13\zD_a\zD^b\tbeta_b$ 
are similarly isolated, 
\beqn
&&\tD_b\tD^b\tbeta_a +\frac13\tD_a\tD_b\tbeta^b 
=\zLap\tbeta_a + \frac13\zD_a \zD^b\tbeta_b
\nonumber\\
&&
+ h^{bc}\zD_b\zD_c\tbeta_a
-\tgamma^{bc}\zD_b(C^d_{ca}\tbeta_d) 
-\tgamma^{bc}C^d_{bc}\tD_d\tbeta_a 
\nonumber\\
&&
-\tgamma^{bc}C^d_{ba}\tD_c\tbeta_d
+ \frac13\zD_a(h^{bc}\zD_b\tbeta_c -\tgamma^{bc}C^d_{bc}\tbeta_d ).  \ \ 
\label{eq:betaterms}
\eeqn
We keep $\tD_a$ instead of replacing it by $\zD_a$ and a connection $C^c_{ab}$
in a couple of terms in the Eq.~(\ref{eq:betaterms}), to shorten the 
equation.
A decomposition proposed by Shibata, 
\beq
\tbeta_a = G_a+\frac18\zD_a(B-x^bG_b), \ \ \ \mbox{where} \ \ \zD_ax^b=\dl_a^b,  
\label{eq:shideco}
\eeq
is substituted in the expression for the flat operator 
$\dis \zLap \tbeta_a +\frac13\zD_a\zD^b\tbeta_b$, 
\beq
\zLap\tbeta_a + \frac13\zD_a \zD^b\tbeta_b 
\,=\, \zLap G_a+ \frac16\zD_a(\zLap B-x^b\zLap G_b), 
\eeq
to obtain elliptic equations that are solved simultaneously,  
\beqn
\zLap G_a &=& \SMom_a, 
\label{eq:shiftG}
\\
\zLap B &=& x^a \SMom_a,   
\label{eq:shiftB}
\eeqn
where the source $\SMom_a$ is written 
\beqn
\SMom_a  
&:=&
-h^{bc}\zD_b\zD_c\tbeta_a
+\tgamma^{bc}\zD_b(C^d_{ca}\tbeta_d)
+\tgamma^{bc}C^d_{bc}\tD_d\tbeta_a 
\nonumber\\
&&
+\tgamma^{bc}C^d_{ba}\tD_c\tbeta_d
- \frac13\zD_a(h^{bc}\zD_b\tbeta_c -\tgamma^{bc}C^d_{bc}\tbeta_d )
\nonumber\\
&&
-\ttR_{ab}\tbeta^b
-\Omega\tD^b \tilde{L}_\phi\tgmabd
-2\alpha\tA_a\!{}^b \frac{\alpha}{\psi^6}
\tD_b \left(\frac{\psi^6}{\alpha}\right)
\nonumber\\
&&
+\frac43\,\alpha\tD_a K
+16\pi \alpha j_a  .
\label{eq:shiftG_source}
\eeqn
A term $\tD^b \tilde{L}_\phi\tgmabd$ 
is computed from 
\beqn
\tD^b \tilde{L}_\phi\tgmabd
&=& 
\tgamma^{bc}\zD_c \tilde{L}_\phi \tgmabd
- C^c_{ba}\tgamma^{bd} \tilde{L}_\phi\tgamma_{cd}
\nonumber\\
&+& \zD_c\tgamma^{cb}\tilde{L}_\phi\tgmabd 
+ C^d_{dc}\tgamma^{cb}\tilde{L}_\phi\tgmabd, 
\eeqn
which is dropped when the WL formulation 
is used (see, Sec.\ref{sec:WL_HS}).

\subsubsection{Spatial trace part of Einstein's equation}

The spatial trace of Einstein's equation is 
combined with the Hamiltonian constraint, 
\beqn
&&\!\!\!\!\!\!\!\!\!
(\Gabd-8\pi\Tabd)(\gamma^{\alpha\beta}+\frac12 n^\alpha n^\beta) 
\nonumber\\
&=&
\,-\, \frac14\tR \,+\,\frac2{\alpha}D^aD_a\alpha  \,+\,
2\Lie_n K 
\nonumber\\
&&
-\, \frac14(K^2+7K_{ab}K^{ab})\,-\,4\pi (\rhoH+2S)=0, \qquad
\eeqn
and it is solved for the combination $\alpha\psi$.  
Using a relation, 
\beqn
&&\!\!\!\!\!\!\!\!\!\!\!
\,-\, \frac14\tR \,+\,\frac2{\alpha}D^aD_a\alpha 
\nonumber\\
&&
\,=\,
\frac2{\alpha\psi^5}\left[\tD^a\tD_a(\alpha\psi)
\,-\,\frac{\alpha\psi}{8}\,\ttR
\right], 
\eeqn
and applying helical symmetry, 
the above equation is rewritten 
\beqn
&&\!\!\!\!\!\!\!\!\!
(\Gabd-8\pi\Tabd)(\gamma^{\alpha\beta}+\frac12 n^\alpha n^\beta)
\nonumber \\
&=&
\frac2{\alpha\psi^5}\left[
\tD^a\tD_a(\alpha\psi)
\,-\,\frac{\alpha\psi}{8}\,\ttR
\,-\,\psi^5\Lie_\omega K 
\right.
\nonumber \\
&&
\left.
\,-\,\alpha\psi^5\left(\frac78 \tA_{ab}\tA^{ab}+\frac5{12}K^2\right)
\,-\,2\pi \,\alpha\psi^5(\rhoH+2S)
\right]
\nonumber \\
&=& 0.
\eeqn
Isolating the flat part $\zLap(\alpha\psi)$, an elliptic equation 
is derived 
\beq
\zLap (\alpha\psi) \,=\, \Str, 
\label{eq:sptrace}
\eeq
where the source $\Str$ is written 
\beqn
\Str 
&:=&
\,-\, h^{ab} \zD_a\zD_b(\alpha\psi) 
\,+\, \tgmabu C^c_{ab}\zD_c(\alpha\psi)
\,+\, \frac{\alpha\psi}{8}\,\ttR
\nonumber \\
&&
\,+\, \psi^5\Lie_\omega K 
\,+\, \alpha\psi^5\left(\frac78 \tA_{ab}\tA^{ab}+\frac5{12}K^2\right)
\nonumber \\
&&
\,+\, 2\pi \,\alpha\psi^5(\rhoH+2S).  
\eeqn

\subsubsection{Spatial tracefree part of Einstein's equation}

The projection of Einstein's equation to the initial hypersurface 
$\Sigma_t$ is written 
\beqn
&&\!\!\!\!\!\!\!\!\!\!\!
(\Gabd -8\pi\Tabd)\gmaa\gmbb  
\nonumber\\
&=&
-\Lie_n \Kabd + \gamma_{ab}\Lie_n K
+ \tR_{ab} - \frac12\gmabd\, \tR  
\nonumber\\
&&
+ K\Kabd-2K_{ac}K_b{}^c\nonumber 
-\frac12\gmabd\left(K^2 +K_{cd}K^{cd}\right)
\nonumber\\
&&
- \frac1{\alpha}(D_aD_b\alpha -\gamma_{ab}D^cD_c\alpha)
-8\pi \Sabd. 
\label{eq:Gab}
\eeqn
The equation to solve for the non-conformal part of the spatial 
metric $h_{ab}$ is derived from the tracefree part of the above 
equation (\ref{eq:Gab}).  
The tracefree operation eliminates 
terms proportional to $\gmabd$.  Applying helical symmetry,
(\ref{eq:Liengmab}) and (\ref{eq:LienKab}), the
tracefree part of Eq.~(\ref{eq:Gab}) is written 
\beq
(\Gabd-8\pi\Tabd)(\gmaa\gmbb-\frac13\gmabd\gamma^{\alpha\beta}) 
\,=\, {\cal E}_{ab}^{\rm TF} \,=\,0, 
\label{eq:tracefree1}
\eeq
where ${\cal E}_{ab}$ is defined by 
\beqn
{\cal E}_{ab}
&:=& \frac1{\alpha}\Lie_\omega \Kabd + \tR_{ab} 
-\frac1{\alpha}D_aD_b\alpha
\nonumber\\
&&
+ K\Kabd -2 K_{ac}K_b{}^c 
-8\pi S_{ab}, 
\label{eq:Eab}
\eeqn
and ${\cal E}_{ab}^{\rm TF}$ is its trace free part
\beq
{\cal E}_{ab}^{\rm TF} 
\,:=\, 
\Big(\gamma_a{}^c\gamma_b{}^d-\frac13\gmabd\gmcdu\Big)
{\cal E}_{cd}
\,=\,
\Big(\tgamma_a{}^c\tgamma_b{}^d-\frac13\tgmabd\tgmcdu\Big)
{\cal E}_{cd}.  
\eeq
The tracefree part of the tensors are also denoted by 
subscripts TF, hereafter.   
We further eliminate terms proportional to $\gmabd$ remaining 
in this expression for ${\cal E}_{ab}$ later in this section.  

We derive two different equations to solve for $h_{ab}$.  
One is an elliptic equation in which $\zLap h_{ab}$ 
is separated from $\tR_{ab}$ as in Eq.~(\ref{eq:laphab});  
it is used for both the WL/NHS formulation.  
The other is for the NHS formulation in which an operator 
$(\zLap - \Omega^2 \pa^2_\phi) h_{ab}$ is separated.  
The $\phi$ derivative term in this operator is separated 
from a term $\frac1\alpha\Lie_\omega \Kabd$, which 
is derived by applying helical symmetry, 
(\ref{eq:Liengmab}) and (\ref{eq:LienKab}), to 
the time derivatives. 
For the former equation, the above ${\cal E}_{ab}$ 
is rewritten 
\beqn
{\cal E}_{ab}
&=& 
- \frac12 \zLap h_{ab} 
+ \Rd_{ab}
+ \Rnl_{ab}
+ \ttR^\psi_{ab} - \frac1{\alpha}D_aD_b\alpha 
\nonumber\\
&&\!\!\!\!\!\!\!
+ K\Kabd -2 K_{ac}K_b{}^c 
+ \frac1\alpha\Lie_\omega \Kabd
- 8\pi S_{ab}, 
\label{eq:EabLap}
\eeqn
and for the latter, 
\beqn
{\cal E}_{ab}
&=& 
- \frac12 \left(\zLap \,-\,\Omega^2\pa^2_{\phi}\right)h_{ab} 
+ \Rd_{ab}
+ \Rnl_{ab}
\nonumber\\
&&
+ \ttR^\psi_{ab} - \frac1{\alpha}D_aD_b\alpha 
+ K\Kabd -2 K_{ac}K_b{}^c 
\nonumber\\
&&
+ \frac1\alpha\Lie_\omega \Kabd
- \frac12 \Omega^2\pa^2_{\phi}h_{ab}
- 8\pi S_{ab}. 
\label{eq:EabHelm}
\eeqn

Terms proportional to $\gmabd$ in Eqs.(\ref{eq:EabLap})
and (\ref{eq:EabHelm}) are now eliminated further
to simplify the equations.  
Introducing barred quantities, 
\beqn
\tbR_{ab}^{\psi} 
&=&-\frac2\psi\tD_a\tD_b\psi+\frac6{\psi^2}\tD_a\psi\tD_b\psi,
\\
\bD_a\bD_b\alpha 
&=&
\zD_a\zD_b\alpha -C^c_{ab}\zD_c\alpha
\nonumber\\
&&-\frac2\psi(\zD_a\alpha\zD_b\psi+\zD_b\alpha\zD_a\psi),
\eeqn
their combination becomes 
\beqn
&&\!\!\!\!\!\!\!\!\!\!\!\!\!\!\!
\tbR_{ab}^{\psi}-\frac1\alpha\bD_a\bD_b\alpha
\nonumber\\
&=&-\,\frac1{\alpha\psi^2} \zD_a\zD_b(\alpha\psi^2)
\,+\,\frac1{\alpha\psi^2} C^c_{ab}\zD_c(\alpha\psi^2)
\nonumber\\
&&\,+\, \frac4{\alpha\psi^2}
\left[\zD_a(\alpha\psi)\zD_b\psi+\zD_b(\alpha\psi)\zD_a\psi\right], 
\eeqn
which satisfies 
\beq
\Big(\ttR_{ab}^{\psi}-\frac1\alpha D_a D_b\alpha\Big)^{\rm TF}
\,=\,
\Big(\tbR_{ab}^{\psi}-\frac1\alpha\bD_a\bD_b\alpha\Big)^{\rm TF}. 
\eeq
Next, substituting $\Kabd = \Aabd + \frac13 \gmabd K$ to terms 
relating to $\Kabd$, their tracefree part satisfies
\beqn
&&
\Big(K K_{ab} -2 K_{ac}K_b{}^c
+ \frac1\alpha\Lie_\omega K_{ab}\Big)^{\rm TF}
\nonumber\\
&&\,=\,
\Big(\frac13 K A_{ab} -2 A_{ac}A_b{}^c
+ \frac1{\alpha}\Lie_\omega A_{ab}\Big)^{\rm TF}.
\eeqn
For the matter source term, 
\beq
S_{ab} =\Tabd\gmaa\gmbb = (\epsilon+p)u_a u_b \,+\, \gmabd p, 
\eeq
where $u_a := \gmaa u_\alpha$.   
We also introduce a barred quantity 
\beq
\bar S_{ab} := \rho{h} u_a u_b, 
\eeq
that satisfies $S_{ab}^{\rm TF}=\bar S_{ab}^{\rm TF}$, 
where $h=(\epsilon+p)/\rho$ is used.  
The tracefree operation to the operator 
$(\zLap \,-\,\Omega^2\pa^2_{\phi}\, )h_{cd} $
is written 
\beqn
&&
- \frac12 
\left(\tgamma_a{}^c\tgamma_b{}^d-\frac13\tgamma_{ab}\tgamma^{cd}\right)
\left(\zLap \,-\,\Omega^2\pa^2_{\phi}\right)h_{cd} 
\nonumber\\
&=& 
-\frac12\left[\left(\zLap \,-\,\Omega^2\pa^2_{\phi}\right)h_{ab} 
\right.
\nonumber\\
&&\left.
\,+\,\frac13\tgmabd\zD^e h^{cd}\,\zD_e h_{cd} 
\,-\,\frac13\tgmabd\Omega^2\pa_{\phi} h^{cd}\pa_{\phi} h_{cd}
\right], \qquad
\eeqn
where relations 
$\tgamma^{cd}\,\zD_e h_{cd}=\tgamma^{cd}\,\pa_\phi h_{cd}=0$ 
implied by $\tgamma = f$ is used. 
The same operation to the Laplacian is written 
similarly as above, but without $\pa_\phi$ terms.

Finally, 
the trace free part ${\cal E}_{ab}^{\rm TF}=0$ results in the following 
elliptic equation, 
\beq
\zLap h_{ab} = {\cal S}_{ab}
\label{eq:WLtrfree}
\eeq
where the source ${\cal S}_{ab}$ is defined by 
\beq
{\cal S}_{ab} :=
2\bar{{\cal E}}_{ab}^{\rm TF}
-\frac13 \tgmabd \zD^e h^{cd}\zD_e h_{cd}, 
\eeq
and $\bar{{\cal E}}_{ab}^{\rm TF}$ is a tracefree 
part of $\bar{{\cal E}}_{ab}$, which is 
written using the rescaled $\tAabd$,
\beqn
\bar{{\cal E}}_{ab}
&:=&
\Rd_{ab} + \Rnl_{ab} + \tbR_{ab}^{\psi}-\frac1\alpha\bD_a\bD_b\alpha
\nonumber\\
&&
+ \frac13 \psi^4 K\tAabd - 2 \psi^4 \tA_{ac}\tA_b{}^c
\nonumber\\
&&
+ \frac1\alpha\Lie_\omega (\psi^4\tAabd)
-8\pi \bar S_{ab}. \qquad
\label{eq:WLEabA}
\eeqn
For the equation with the operator $\zLap-\Omega^2\pa^2_\phi$, 
it is written 
\beq
\left(\zLap-\Omega^2\pa^2_\phi\right) h_{ab} = {\cal S}_{ab}
\label{eq:NHStrfree}
\eeq
with
\beq
{\cal S}_{ab} :=
2\bar{{\cal E}}_{ab}^{\rm TF}
-\frac13 \tgmabd \zD^e h^{cd}\zD_e h_{cd}
+\frac13 \tgmabd \Omega^2\pa_\phi h^{cd}\pa_\phi h_{cd}. 
\eeq
Using the rescaled $\tAabd$, $\bar{{\cal E}}_{ab}$ is defined by
\beqn
\bar{{\cal E}}_{ab}
&:=&
\Rd_{ab} + \Rnl_{ab} 
+ \tbR_{ab}^{\psi}-\frac1\alpha\bD_a\bD_b\alpha
\nonumber\\
&&
+\, \frac13\psi^4 K\tAabd - 2 \psi^4\tA_{ac}\tA_b{}^c
\nonumber\\
&&
+\, \frac1\alpha\Lie_\omega (\psi^4\tAabd)
- \frac12 \Omega^2\pa^2_{\phi}h_{ab} 
-8\pi \bar S_{ab}, \qquad
\label{eq:NHSEabA}
\eeqn
where a difference from (\ref{eq:WLEabA}) is a term in the 
last line.

\subsubsection{Matter source terms}
\label{sec:source}

In the above, the matter source terms, $\rhoH$, $j_a$, $S$ and  $S_{ab}$, 
that appear in the field equations are obtained from the stress energy tensor.  
We write the projection of the stress energy tensor in terms of the 
fluid variables and metric potentials.  
The 4-velocity for irrotational flow $u^\alpha = u^t (k^\alpha+v^\alpha)$ 
is decomposed with respect to the foliation $\Sigma_t$ as 
\begin{eqnarray}
u^\alpha n_\alpha &=& -\alpha u^t
\\
u^\alpha \gamma_{\alpha a} &=& u_a = \frac1{h}D_a \Phi = \frac1{h}\zD_a \Phi, 
\end{eqnarray}
where the velocity potential $\Phi$ is introduced by 
$hu_\alpha = \na_\alpha \Phi$,

Using these relations, the matter source terms of the field equations 
become 
\beqn
\rhoH \,&:=&\, \Tabd n^\alpha n^\beta 
\,=\, h\rho  (\alpha u^t)^2 - p,
\\
j_a \,&:=&\, -\Tabd \gmaa n^\beta
\,=\, \rho \alpha u^t \zD_a \Phi,
\\
S \,&:=&\, \Tabd \gamma^{\albe}
\,=\, h \rho \big[(\alpha u^t)^2 - 1\big] +3\,p, 
\\
S_{ab} &:=&\Tabd\gmaa\gmbb = \frac\rho{h}\zD_a\Phi\,\zD_b\Phi+ p\gmabd , 
\eeqn
or with a barred quantity, 
\beq
\bar S_{ab} = \frac\rho{h}\zD_a\Phi\,\zD_b\Phi.
\eeq

\subsection{Equations for irrotational fluid}
\label{sec:irrotfluid}

Following Sec.\ref{sec:matter} and \ref{sec:pEOS}, a set of 
equations used in our codes to solve for the matter variables 
are derived.  
As independent variables, we choose 
the relativistic enthalpy per baryon mass, 
the time component of the 4-velocity, and the 
velocity potential, $\{h,u^t,\Phi\}$.  
For the first two variables, 
the first integral Eq.~(\ref{eq:firstint}) and 
the normalization of the 
4-velocity $u_\alpha u^\alpha=-1$ are solved.  
Using a relation derived from Eqs.(\ref{eq:vpot}) 
and (\ref{eq:udecomp}), 
\beq
v_a + \omega_a = \frac1{hu^t}D_a\Phi, 
\label{eq:spatialvpot}
\eeq
these equations are rewritten, 
\beqn
h &=& 
\left[
\frac1{\alpha^2}
\left({\cal E}+\omega^a D_a\Phi\right)^2
-D_a\Phi D^a\Phi
\right]^{1/2}, 
\label{eq:h}
\\
u^t &=& \frac1{\alpha^2 h}({\cal E} + \omega^a D_a \Phi), 
\label{eq:ut}
\eeqn
where the first one is from $u_\alpha u^\alpha=-1$, 
and the second from Eq.~(\ref{eq:firstint}).  

An equation to calculate the velocity potential $\Phi$ is derived 
from the rest mass conservation law, Eq.~(\ref{eq:masscon_eqL}), 
\beqn
\frac{1}{\sqrt{-g}}\Lie_u (\rho\sqrt{-g})
&=& \frac{1}{\alpha\sqrt{\gamma}}\Lie_v (\rho u^t \alpha\sqrt{\gamma}) 
\nonumber\\
&=& \frac{1}{\alpha}D_a(\alpha \rho  u^t v^a) =0.  
\eeqn
Substituting Eq.~(\ref{eq:spatialvpot}) in the above relation, 
we have an elliptic equation for $\Phi$, 
\beq
D^a D_a\Phi = D_a(hu^t\omega^a)
- (D_a\Phi - hu^t\omega_a)\frac{h}{\alpha\rho}D^a\frac{\alpha\rho}{h}.
\label{eq:LapPhi}
\eeq
This equation is solved with Neumann boundary condition 
to impose the fluid 4-velocity $u^\alpha$ to follow 
the surface of the star.
The surface is defined by the vanishing pressure $p=0$, which 
coincide with the $h=1$ surface in our EOS (see, Sec \ref{sec:pEOS}).  
Hence, the boundary condition is written 
\beq
u^\alpha \na_\alpha h =0  \ \ \mbox{ at }\ \  h=1.  
\eeq
and, using $\Lie_k h = 0$ and Eq.~(\ref{eq:spatialvpot}), 
Neumann boundary condition 
for the potential $\Phi$ is rewritten, 
\beq
(D^a\Phi - hu^t\omega^a)D_a h =0.
\eeq
where $D_a h$ is normal to the stellar surface.  

Finally we rewrite the above set of equations 
for the helically symmetric irrotational flow 
using the flat derivative $\zD_a$;
\beqn
& \dis
h \,=\,
\left[
\frac1{\alpha^2}
\left({\cal E}+\tomega^a \zD_a\Phi\right)^2
-\frac1{\psi^4}\tgmabu\zD_a\Phi \zD_b\Phi
\right]^{1/2}, \qquad& 
\label{eq:hzD}
\\
&\dis 
u^t \,=\, \frac1{\alpha^2 h}({\cal E} + \tomega^a \zD_a \Phi), &
\label{eq:utzD}
\\[2mm]
&\zLap \Phi \,=\, {\cal S}, &
\label{eq:LapPhizD}
\eeqn
where ${\cal S}$ is defined by
\beqn
{\cal S}&=&
-\,h^{ab}\zD_a\zD_b\Phi
\,+\,\tgmabu C^c_{ab}\zD_c\Phi
\,-\,\frac2{\psi}\tgmabu\zD_a\psi \zD_b\Phi
\nonumber\\
&&
\,+\,\frac1{\psi^2}\,\tomega^a\zD_a(hu^t\psi^6) 
\,+\,\psi^4 h u^t \tD_a\tomega^a
\nonumber\\
&&
\,-\,(\tgmabu\zD_b\Phi - \psi^4hu^t\tomega^a)
\frac{h}{\alpha\rho}\zD_a\frac{\alpha\rho}{h}, 
\label{eq:LapPhisource}
\eeqn
and, for $\tgamma=f$, 
$\tD_a\tomega^a=\zD_a\tomega^a=\zD_a\tbeta^a$.

\section{Self-consistent field iteration scheme}
\label{secA:nume}

\subsection{Elliptic equation solver}

As mentioned in Sec.\ref{sec:coord}, components of the metric 
are computed on a spherical-coordinate grid whose origin is placed 
at the center of mass.  The momentum constraints and 
the tracefree part of Einstein's equation are a 
spatial vector and a tensor equation, respectively, 
and it would be natural to write the equations 
in components along the spherical coordinates \cite{USE00}.  
It is simpler, however, to solve these equations for 
cartesian components, yet on the spherical coordinates, 
because each cartesian component satisfies a field equation
whose principal part is the same as that of a scalar equation. 

For the spatial tracefree part of Einstein's 
equation solved for the non-conformally flat part $h_{ab}$, 
writing the principal part ${\cal L} := \zLap$ or 
$\zLap-\Omega^2 \pa^2_\phi$, the field equations become
\beq
{\cal L} h_{ab} = {\cal S}_{ab}.  
\eeq
Expanding each cartesian component of $h_{ab}$ in scalar multipoles, 
the equation with the operator 
$\zLap-\Omega^2 \pa^2_\phi$ becomes a Helmholtz equation for each 
mode, 
\beq
(\zLap+m^2\Omega^2) h_{ab}^{\ell m}Y_{\ell m}
={\cal S}_{ab}^{\ell m}Y_{\ell m}.  
\eeq
Hence these elliptic equations are integrated using Green's formula, 
\beqn
h_{ab}(x)&=& -\frac1{4\pi}\int_{V} G(x,x'){\cal S}_{ab}(x') d^{3}x' 
\nonumber \\
&&
+ \frac{1}{4\pi} \int_{\pa V} \left[G(x,x')\zD'^c h_{ab}(x')\right.
\nonumber \\
&&\qquad\qquad
\left. - h_{ab}(x')\zD'^c G(x,x') \right]dS'_c.  
\label{eq:GreenIde}
\eeqn
where $x$ and $x'$ are positions, $x,x'\in V \subseteq \Sigma_t$, 
and the Green function $G(x,x')$ satisfies
\beq
{\cal L} G(x,x') = -4\pi\delta(x-x').  
\eeq
We choose the Green function $G(x,x')$ without boundary 
for the BNS calculations.  

For the Laplace operator, ${\cal L} =\zLap$, 
a multipole expansion of $G(x,x')$ 
in associated Legendre functions
on the spherical coordinate is written 
\beqn
&&
G(x,x')=\frac{1}{\left|{x}-{x'}\right|}\,=\, 
\sum_{\ell=0}^\infty g_\ell(r,r') \sum_{m=0}^\ell \epsilon_m \,
\frac{(\ell-m)!}{(\ell+m)!}
\nonumber\\
&&\qquad\quad
\times
P_\ell^{~m}(\cos\theta)\,P_\ell^{~m}(\cos\theta')
\cos m(\varphi-\varphi'), 
\eeqn
where the radial Green function $g_\ell(r,r')$ becomes 
\beq
g_\ell(r,r')=\frac{r_<^\ell}{r_>^{\ell+1}}, 
\eeq
with 
$
r_> := \sup\{r,r'\}, \ r_< := \inf\{r,r'\}, 
$
and the coefficients $\epsilon_m$ are equal to $\epsilon_0 = 1$ for $m=0$, 
and $\epsilon_m = 2$ for $m\ge 1$.

For the case with the Helmholtz operator, ${\cal L} =\zLap+m^2\Omega^2 $, 
we choose the Green function for the half-retarded + half-advanced 
field \cite{YBRUF06}, 
\beqn
&&
G(x,x')\,=\, 
\sum_{\ell=0}^\infty \sum_{m=0}^\ell g_{\ell m}(r,r') \epsilon_m \,
\frac{(\ell-m)!}{(\ell+m)!}
\nonumber\\
&&\qquad\quad
\times
P_\ell^{~m}(\cos\theta)\,P_\ell^{~m}(\cos\theta')
\cos m(\varphi-\varphi'), \quad
\eeqn
where the radial Green function $g_{\ell m}(r,r')$ is constructed 
from the spherical Bessel function of the first and second kinds 
$j_\ell(x)$ and $n_\ell(x)$, 
\beqn
&&
g_{\ell m}(r,r')\,=\,\left\{ \begin{array}{l}
\displaystyle \frac{r_<^\ell}{r_>^{\ell+1}},  \qquad\ \mbox{for }\ m=0, 
\\[4mm]
-m\Omega \,(2\ell+1)\, j_\ell(m\Omega\, r_<)\, n_\ell(m\Omega\, r_>), 
           \end{array} \right.
\\[1mm]
&&\qquad\qquad\qquad\qquad\qquad\qquad
 \mbox{for }\ m\geq 1.
\nonumber
\eeqn

\subsection{Summary for iteration scheme}
\label{secA:numeiter}

Eq.~(\ref{eq:GreenIde}) is used as an elliptic equation solver  
for the field variables $\{\alpha,\beta_a,\psi,h_{ab}\}$.  
In the code, the elliptic solver is used to compute  
the combination $\alpha \psi$ from Eq.~(\ref{eq:sptrace}); 
to compute the potentials (\ref{eq:shideco}) of the shift vector 
$\tbeta_a$ from Eqs.~(\ref{eq:shiftG}) and (\ref{eq:shiftB}); 
and to compute the gauge potentials (\ref{eq:DgaugeBY}) from 
Eq.~(\ref{eq:Dgauge_ellip}).  

For the fluid variables, $\{h, u^t, \Phi \}$ are found from 
Eqs.(\ref{eq:hzD}), (\ref{eq:utzD}), and (\ref{eq:LapPhizD}), 
respectively.  A detailed description of a method to solve 
Eq.~(\ref{eq:LapPhizD}) is found in \cite{USE00}.  
As we use the surface-fitted coordinates to 
calculate neutron stars, the surface $R(\theta_f,\phi_f)$ becomes 
an additional variable.  A stellar surface is defined by the 
pressure $p=0$, and, instead, it is located by a condition 
$q=p/\rho=0$ in the code.  

A solution is specified by two parameters for an equal mass binary, 
which we take to be the orbital angular momentum and the injection energy, 
$\{\Omega, \cal E\}$.  We introduce one more parameter $R_0$ 
to normalize the radial coordinate, where $R_0$ is 
half the coordinate diameter of a neutron star along 
the $(\theta_f,\phi_f)=(\pi/2,0)$ line.  
These parameters are calculated from the conditions 
$R(\pi/2,0)/R_0=1$ and $R(\pi/2,\pi)/R_0=1$, after 
prescribing a value of a thermodynamic variable at 
a point in a star, for which a central value of $h$
is fixed at $r_f = 0$.  These conditions are applied to 
Eq.~(\ref{eq:hzD}), and solved for the three parameters.  

All these variables are assigned on each grid point, and the parameters 
are calculated from the equations mentioned above 
in each iteration cycle.  If we represent the set of 
fluid and metric variables by $\hat \Psi$, we can describe the 
iteration schematically as follows.  
The variables are are updated from their values at the Nth iteration cycle, 
$\Psi^{({\rm N})}$, to the (N+1)th, $\Psi^{({\rm N+1})}$, 
using softening, in the manner 
\beq
\Psi^{({\rm N+1})}
\,=\,
\lambda \hat \Psi
\,+\,(1-\lambda) \Psi^{({\rm N})},
\eeq
where $\lambda$ is the softening parameter, 
chosen to be in the range $0.1$ to $0.3$ to accelerate convergence. 
For a criteria to determine convergence, a relative
difference of successive cycles 
\beq
\frac{2|\Psi^{({\rm N+1})} - \Psi^{({\rm N})}|}
{|\Psi^{({\rm N+1})}| + |\Psi^{({\rm N})}|} < \delta
\eeq
is used, with $\delta = 10^{-6}$ 
in the present calculations.

\begin{table}
\begin{tabular}{ccccccccccc}
\hline
$d/R_0$ & $2d/M$ & $R_0/M$ & $\Omega M$ & $\Madm$ & $J/M^2$  \\
\hline
$ 1.3125 $ & $  9.2784     $ & $ 7.0693 $ & $ 0.030149 $ & $ 2.67191 $ & $ 0.96637     $ \\ 
$ 1.3438 $ & $  9.4190     $ & $ 7.0095 $ & $ 0.029649 $ & $ 2.67206 $ & $ 0.96819     $ \\ 
$ 1.3750 $ & $  9.5654     $ & $ 6.9566 $ & $ 0.029126 $ & $ 2.67223 $ & $ 0.97048     $ \\ 
$ 1.4375 $ & $  9.8715     $ & $ 6.8671 $ & $ 0.028019 $ & $ 2.67265 $ & $ 0.97582     $ \\ 
$ 1.5000 $ & $  10.192     $ & $ 6.7948 $ & $ 0.026901 $ & $ 2.67311 $ & $ 0.98279     $ \\ 
$ 1.6250 $ & $  10.873     $ & $ 6.6912 $ & $ 0.024693 $ & $ 2.67422 $ & $ 0.99905     $ \\ 
$ 1.7500 $ & $  11.590     $ & $ 6.6229 $ & $ 0.022648 $ & $ 2.67537 $ & $  1.0178     $ \\ 
$ 1.8750 $ & $  12.330     $ & $ 6.5759 $ & $ 0.020794 $ & $ 2.67648 $ & $  1.0375     $ \\ 
$ 2.0000 $ & $  13.090     $ & $ 6.5451 $ & $ 0.019118 $ & $ 2.67756 $ & $  1.0572     $ \\ 
$ 2.5000 $ & $  16.234     $ & $ 6.4936 $ & $ 0.014113 $ & $ 2.68130 $ & $  1.1415     $ \\ 
$ 3.0000 $ & $  19.455     $ & $ 6.4852 $ & $ 0.010889 $ & $ 2.68404 $ & $  1.2229     $ \\ 
\hline
\end{tabular}
\caption{Solution sequence for the EOS 2H.  
}
\label{tab:seq_2H}
%
\vskip 5mm
%
\begin{tabular}{ccccccccccc}
\hline
$d/R_0$ & $2d/M$ & $R_0/M$ & $\Omega M$ & $\Madm$ & $J/M^2$  \\
\hline
$ 1.3750 $ & $  6.8380     $ & $ 4.9731 $ & $ 0.045877 $ & $ 2.66507 $ & $ 0.89082     $ \\ 
$ 1.4062 $ & $  6.9403     $ & $ 4.9353 $ & $ 0.045086 $ & $ 2.66521 $ & $ 0.89216     $ \\ 
$ 1.4375 $ & $  7.0511     $ & $ 4.9051 $ & $ 0.044231 $ & $ 2.66536 $ & $ 0.89378     $ \\ 
$ 1.5000 $ & $  7.2800     $ & $ 4.8533 $ & $ 0.042526 $ & $ 2.66580 $ & $ 0.89812     $ \\ 
$ 1.6250 $ & $  7.7600     $ & $ 4.7754 $ & $ 0.039177 $ & $ 2.66688 $ & $ 0.90829     $ \\ 
$ 1.7500 $ & $  8.2681     $ & $ 4.7246 $ & $ 0.036025 $ & $ 2.66810 $ & $ 0.92072     $ \\ 
$ 1.8750 $ & $  8.7975     $ & $ 4.6920 $ & $ 0.033145 $ & $ 2.66932 $ & $ 0.93485     $ \\ 
$ 2.0000 $ & $  9.3425     $ & $ 4.6712 $ & $ 0.030526 $ & $ 2.67057 $ & $ 0.94926     $ \\ 
$ 2.5000 $ & $  11.606     $ & $ 4.6423 $ & $ 0.022621 $ & $ 2.67511 $ & $  1.0132     $ \\ 
$ 3.0000 $ & $  13.931     $ & $ 4.6435 $ & $ 0.017521 $ & $ 2.67860 $ & $  1.0789     $ \\ 
\hline
\end{tabular}
\caption{Solution sequence for the EOS HB.  
}
\label{tab:seq_HB}
%
\vskip 5mm
%
\begin{tabular}{ccccccccccc}
\hline
$d/R_0$ & $2d/M$ & $R_0/M$ & $\Omega M$ & $\Madm$ & $J/M^2$  \\
\hline
$ 1.4375 $ & $  5.5971     $ & $ 3.8936 $ & $ 0.059912 $ & $ 2.66109 $ & $ 0.85642     $ \\ 
$ 1.4688 $ & $  5.6801     $ & $ 3.8673 $ & $ 0.058871 $ & $ 2.66118 $ & $ 0.85733     $ \\ 
$ 1.5000 $ & $  5.7713     $ & $ 3.8475 $ & $ 0.057759 $ & $ 2.66133 $ & $ 0.85840     $ \\ 
$ 1.5313 $ & $  5.8622     $ & $ 3.8284 $ & $ 0.056652 $ & $ 2.66150 $ & $ 0.85959     $ \\ 
$ 1.5625 $ & $  5.9542     $ & $ 3.8107 $ & $ 0.055575 $ & $ 2.66171 $ & $ 0.86099     $ \\ 
$ 1.6250 $ & $  6.1443     $ & $ 3.7811 $ & $ 0.053414 $ & $ 2.66211 $ & $ 0.86424     $ \\ 
$ 1.7500 $ & $  6.5438     $ & $ 3.7393 $ & $ 0.049241 $ & $ 2.66318 $ & $ 0.87225     $ \\ 
$ 1.8750 $ & $  6.9613     $ & $ 3.7127 $ & $ 0.045416 $ & $ 2.66432 $ & $ 0.88246     $ \\ 
$ 2.0000 $ & $  7.3930     $ & $ 3.6965 $ & $ 0.041907 $ & $ 2.66555 $ & $ 0.89311     $ \\ 
$ 2.5000 $ & $  9.1954     $ & $ 3.6782 $ & $ 0.031170 $ & $ 2.67036 $ & $ 0.94283     $ \\ 
$ 3.0000 $ & $  11.050     $ & $ 3.6835 $ & $ 0.024219 $ & $ 2.67433 $ & $ 0.99755     $ \\ 
\hline
\end{tabular}
\caption{Solution sequence for the EOS 2B.  
}
\label{tab:seq_2B}
\end{table}
%
%
\begin{table}
\begin{tabular}{ccccccccccc}
\hline
$d/R_0$ & $2d/M$ & $R_0/M$ & $\Omega M$ & $\Madm$ & $J/M^2$  \\
\hline
$ 1.3750 $ & $  6.7360     $ & $ 4.8989 $ & $ 0.046804 $ & $ 2.66479 $ & $ 0.8804     $ \\ 
$ 1.4375 $ & $  6.9455     $ & $ 4.8316 $ & $ 0.045130 $ & $ 2.66505 $ & $ 0.8987     $ \\ 
$ 1.4687 $ & $  7.0572     $ & $ 4.8049 $ & $ 0.044256 $ & $ 2.66526 $ & $ 0.8983     $ \\ 
$ 1.5000 $ & $  7.1692     $ & $ 4.7795 $ & $ 0.043403 $ & $ 2.66546 $ & $ 0.8998     $ \\ 
$ 1.5625 $ & $  7.4012     $ & $ 4.7367 $ & $ 0.041683 $ & $ 2.66600 $ & $ 0.8952     $ \\ 
$ 1.6250 $ & $  7.6424     $ & $ 4.7030 $ & $ 0.039990 $ & $ 2.66655 $ & $ 0.9093     $ \\ 
$ 1.7500 $ & $  8.1425     $ & $ 4.6529 $ & $ 0.036778 $ & $ 2.66777 $ & $ 0.9109     $ \\ 
$ 1.8750 $ & $  8.6640     $ & $ 4.6208 $ & $ 0.033842 $ & $ 2.66899 $ & $ 0.9397     $ \\ 
$ 2.0000 $ & $  9.2007     $ & $ 4.6004 $ & $ 0.031172 $ & $ 2.67025 $ & $ 0.9413     $ \\ 
$ 2.5000 $ & $  11.431     $ & $ 4.5722 $ & $ 0.023105 $ & $ 2.67481 $ & $  1.082     $ \\ 
$ 3.0000 $ & $  13.721     $ & $ 4.5737 $ & $ 0.017898 $ & $ 2.67834 $ & $  1.031     $ \\ 
\hline
\end{tabular}
\caption{Solution sequence for the EOS SLy.  
}
\label{tab:seq_SLy}
%
\vskip 5mm
%
\begin{tabular}{ccccccccccc}
\hline
$d/R_0$ & $2d/M$ & $R_0/M$ & $\Omega M$ & $\Madm$ & $J/M^2$  \\
\hline
$ 1.6875 $ & $  5.8207     $ & $ 3.4493 $ & $ 0.057394 $ & $ 2.66144 $ & $ 0.85585     $ \\ 
$ 1.7500 $ & $  6.0069     $ & $ 3.4325 $ & $ 0.055113 $ & $ 2.66195 $ & $ 0.85933     $ \\ 
$ 1.8125 $ & $  6.1950     $ & $ 3.4179 $ & $ 0.052972 $ & $ 2.66247 $ & $ 0.86339     $ \\ 
$ 1.8750 $ & $  6.3877     $ & $ 3.4068 $ & $ 0.050902 $ & $ 2.66302 $ & $ 0.86770     $ \\ 
$ 1.9375 $ & $  6.5845     $ & $ 3.3985 $ & $ 0.048909 $ & $ 2.66363 $ & $ 0.87233     $ \\ 
$ 2.0000 $ & $  6.7839     $ & $ 3.3920 $ & $ 0.047008 $ & $ 2.66423 $ & $ 0.87706     $ \\ 
$ 2.5000 $ & $  8.4399     $ & $ 3.3760 $ & $ 0.035057 $ & $ 2.66900 $ & $ 0.92198     $ \\ 
$ 3.0000 $ & $  10.148     $ & $ 3.3826 $ & $ 0.027269 $ & $ 2.67308 $ & $ 0.97231     $ \\ 
\hline
\end{tabular}
\caption{Solution sequence for the EOS APR1.  
}
\label{tab:seq_APR1}
%
\vskip 5mm
%
\begin{tabular}{ccccccccccc}
\hline
$d/R_0$ & $2d/M$ & $R_0/M$ & $\Omega M$ & $\Madm$ & $J/M^2$  \\
\hline
$ 1.4375 $ & $  6.3451     $ & $ 4.4140 $ & $ 0.050851 $ & $ 2.66331 $ & $ 0.87458     $ \\ 
$ 1.4688 $ & $  6.4477     $ & $ 4.3899 $ & $ 0.049882 $ & $ 2.66346 $ & $ 0.87609     $ \\ 
$ 1.5000 $ & $  6.5504     $ & $ 4.3669 $ & $ 0.048919 $ & $ 2.66364 $ & $ 0.87782     $ \\ 
$ 1.5313 $ & $  6.6534     $ & $ 4.3451 $ & $ 0.047977 $ & $ 2.66387 $ & $ 0.87955     $ \\ 
$ 1.5625 $ & $  6.7597     $ & $ 4.3262 $ & $ 0.047029 $ & $ 2.66415 $ & $ 0.88156     $ \\ 
$ 1.6250 $ & $  6.9799     $ & $ 4.2953 $ & $ 0.045139 $ & $ 2.66467 $ & $ 0.88618     $ \\ 
$ 1.7500 $ & $  7.4359     $ & $ 4.2491 $ & $ 0.041551 $ & $ 2.66585 $ & $ 0.89672     $ \\ 
$ 1.8750 $ & $  7.9127     $ & $ 4.2201 $ & $ 0.038262 $ & $ 2.66709 $ & $ 0.90918     $ \\ 
$ 2.0000 $ & $  8.4036     $ & $ 4.2018 $ & $ 0.035261 $ & $ 2.66835 $ & $ 0.92183     $ \\ 
$ 2.5000 $ & $  10.447     $ & $ 4.1786 $ & $ 0.026166 $ & $ 2.67304 $ & $ 0.97944     $ \\ 
$ 3.0000 $ & $  12.547     $ & $ 4.1822 $ & $ 0.020291 $ & $ 2.67677 $ & $  1.0401     $ \\ 
\hline
\end{tabular}
\caption{Solution sequence for the EOS FPS.  
}
\label{tab:seq_FPS}
\end{table}
%
%
\begin{table}
\begin{tabular}{ccccccccccc}
\hline
$d/R_0$ & $2d/M$ & $R_0/M$ & $\Omega M$ & $\Madm$ & $J/M^2$  \\
\hline
$ 1.4375 $ & $  7.8579     $ & $ 5.4664 $ & $ 0.038232 $ & $ 2.66765 $ & $ 0.91610     $ \\ 
$ 1.4062 $ & $  7.7372     $ & $ 5.5020 $ & $ 0.038958 $ & $ 2.66749 $ & $ 0.91413     $ \\ 
$ 1.4688 $ & $  7.9817     $ & $ 5.4344 $ & $ 0.037499 $ & $ 2.66786 $ & $ 0.91847     $ \\ 
$ 1.5000 $ & $  8.1086     $ & $ 5.4058 $ & $ 0.036772 $ & $ 2.66810 $ & $ 0.92130     $ \\ 
$ 1.5625 $ & $  8.3727     $ & $ 5.3585 $ & $ 0.035291 $ & $ 2.66867 $ & $ 0.92711     $ \\ 
$ 1.6875 $ & $  8.9226     $ & $ 5.2875 $ & $ 0.032465 $ & $ 2.66983 $ & $ 0.94095     $ \\ 
$ 1.7500 $ & $  9.2099     $ & $ 5.2628 $ & $ 0.031101 $ & $ 2.67046 $ & $ 0.94804     $ \\ 
$ 1.8750 $ & $  9.7970     $ & $ 5.2250 $ & $ 0.028611 $ & $ 2.67166 $ & $ 0.96421     $ \\ 
$ 2.0000 $ & $  10.402     $ & $ 5.2008 $ & $ 0.026341 $ & $ 2.67286 $ & $ 0.98037     $ \\ 
$ 2.5000 $ & $  12.914     $ & $ 5.1657 $ & $ 0.019492 $ & $ 2.67718 $ & $  1.0507     $ \\ 
$ 3.0000 $ & $  15.493     $ & $ 5.1643 $ & $ 0.015068 $ & $ 2.68040 $ & $  1.1206     $ \\ 
\hline
\end{tabular}
\caption{Solution sequence for the EOS BGN1H1.  
}
\label{tab:seq_BGN1H1}
%
\vskip 5mm
%
\begin{tabular}{ccccccccccc}
\hline
$d/R_0$ & $2d/M$ & $R_0/M$ & $\Omega M$ & $\Madm$ & $J/M^2$  \\
\hline
$ 1.4375 $ & $  6.0667     $ & $ 4.2203 $ & $ 0.053901 $ & $ 2.66264 $ & $ 0.86655     $ \\ 
$ 1.4688 $ & $  6.1649     $ & $ 4.1974 $ & $ 0.052878 $ & $ 2.66279 $ & $ 0.86785     $ \\ 
$ 1.5000 $ & $  6.2628     $ & $ 4.1752 $ & $ 0.051867 $ & $ 2.66297 $ & $ 0.86944     $ \\ 
$ 1.6250 $ & $  6.6732     $ & $ 4.1066 $ & $ 0.047891 $ & $ 2.66396 $ & $ 0.87718     $ \\ 
$ 1.6875 $ & $  6.8881     $ & $ 4.0818 $ & $ 0.045973 $ & $ 2.66451 $ & $ 0.88202     $ \\ 
$ 1.7500 $ & $  7.1090     $ & $ 4.0623 $ & $ 0.044109 $ & $ 2.66512 $ & $ 0.88709     $ \\ 
$ 1.8125 $ & $  7.3353     $ & $ 4.0470 $ & $ 0.042335 $ & $ 2.66576 $ & $ 0.89298     $ \\ 
$ 1.8750 $ & $  7.5659     $ & $ 4.0352 $ & $ 0.040622 $ & $ 2.66635 $ & $ 0.89878     $ \\ 
$ 2.0000 $ & $  8.0365     $ & $ 4.0183 $ & $ 0.037449 $ & $ 2.66765 $ & $ 0.91094     $ \\ 
$ 2.5000 $ & $  9.9949     $ & $ 3.9980 $ & $ 0.027806 $ & $ 2.67237 $ & $ 0.96616     $ \\ 
$ 3.0000 $ & $  12.009     $ & $ 4.0031 $ & $ 0.021559 $ & $ 2.67620 $ & $  1.0244     $ \\ 
\hline
\end{tabular}
\caption{Solution sequence for the EOS ALF3.  
}
\label{tab:seq_ALF3}
\end{table}

\section{Formulas for mass and angular momentums}

Definitions of the quantities shown in tables and figures, which 
characterize a solution of BNS, 
and their expressions used in actual numerical computations, 
are summarized in this Appendix.  

The rest mass is the baryon mass density measured by 
comoving observers integrated over the initial hypersurface, 
and during the inspiral phase of binary neutron star, 
it is considered to be conserved.  
The rest mass of one component of a binary system is 
written $M_0$ and defined by 
\beq
M_{0}
\,:=\, \int_\Sigma \rho\,u^\alpha dS_\alpha 
\,=\, \int_\Sigma \rho u^t \alpha \psi^6\sqrt{\tgamma}d^3x
\eeq
where $dS_\alpha = \na_\alpha t \sqrt{-g} d^3x$, 
and $\sqrt{-g}d^3x=\alpha\psi^6\sqrt{\tgamma}d^3x$
$=\alpha\psi^6 r^2 \sin\theta dr d\theta d\phi$, 
because $\tgamma=f$ is assumed.  
 
In this paper, the mass $M_1$ is used to specify 
an equal mass BNS sequence, and $M = 2 M_1$ 
is used to normalize quantities.  $M_1$ is 
the gravitational mass of a single spherical star 
whose rest mass is equal to the rest mass $M_0$ of 
one neutron star in the binary system of each model
(see Table \ref{tab:EOS}).

The ADM mass $\Madm$ is rewritten using conformal spatial metric,   
\beqn
\Madm
&:=& \frac1{16\pi}\int_\infty 
\left(f^{ac}f^{bd}-f^{ab}f^{cd}\right)
\zD_b\gamma_{cd}\, dS_a
\nonumber\\
&:=& \frac1{16\pi}\int_\infty 
\left(f^{ac}f^{bd}-f^{ab}f^{cd}\right)
\zD_b\tgamma_{cd}\, dS_a
\nonumber\\
&+& \frac1{16\pi}\int_\infty(-2) f^{ab}\zD_b\psi^4 \,dS_a
\nonumber\\
&=& 
- \frac1{2\pi}\int_\infty \zD{}^a\psi \,dS_a, 
\eeqn
where, in the second equality, the first term 
vanishes because of our choice $\tgamma = f$; 
and $\psi\rightarrow 1$ is used in the second term.  
We have calculated approximate values of $\Madm$ using 
this surface integral at the boundary of the computational domain.  
Also, we fit $M_\psi/2r$ to $\psi -1$ near the boundary, 
to ensure a constant $M_\psi \approx \Madm$.  In the tables, 
however, the values of $\Madm$ are calculated from a formula 
in which the above surface integral is converted to a volume 
integral using the Gauss-Stokes lemma.  We apply this 
on the conformal spatial hypersurface, which results in 
a simpler formula; since, at spatial infinity
$\psi\rightarrow 1$, $\tgamma^{ab}\rightarrow f^{ab}$and
$dS_a = \na_a r \sqrt{f}d^2x = \na_a r \sqrt{\tgamma}d^2x=:d\tS_a$, 
we have 
\beqn
\Madm
&=& 
- \frac1{2\pi}\int_\infty \tD{}^a\psi \,d\tS_a, 
\,=\, 
- \frac1{2\pi}\int_\Sigma \tLap\psi \,d\tS, 
\nonumber\\
&=& \frac1{2\pi}\int_\Sigma \left[\,
\,-\,\frac{\psi}{8}\,\ttR
\,+\,\frac18\psi^5\left(\tA_{ab}\tA^{ab}-\frac23 K^2\right)\right.
\nonumber\\
&&\left.\phantom{\frac11}
\,+\,2\pi\psi^5\rhoH \,\right]\sqrt{\tgamma}d^3x.
\eeqn

The Komar mass associated with a timelike 
Killing field $t^\alpha$ is written   
\beqn
\MK
&:=& -\frac1{4\pi}\int_\infty \na^\alpha\, t^\beta \,dS_{\albe}
\,=\,-\frac1{4\pi}\int_\Sigma \Rab t^\beta\,dS_\alpha
\nonumber\\
&=&  \int_\Sigma \left(\,2\Tab - T \gab \,\right)\,t^\beta \,dS_\alpha, 
\nonumber\\
&=& \int_\Sigma\left[\, \alpha \left(\rhoH+S\right) 
-2 j_a \beta^a\,\right] \psi^6\sqrt{\tgamma}d^3x, 
\eeqn
where $dS_\alpha = n_\alpha \sqrt{\gamma}d^3x$ is used.  
To derive this, the global existence of a timelike 
Killing field is assumed.  For the spacetime of WL/NHS formulations, 
no such timelike Killing field exists. Instead, an asymptotic Komar 
mass can be written 
\beqn
\MK
&:=& -\frac1{4\pi}\int_\infty \na^\alpha t^\beta \,dS_{\albe}
\,=\,\frac1{4\pi}\int_\infty D^a \alpha\,dS_a
\nonumber\\
&=&  \frac1{4\pi}\int_\Sigma \Dl \alpha\,d\Sigma
\nonumber\\
&=& \frac1{4\pi}\int_\Sigma\left[\, 
\left(\alpha \tA_{ab}\tA^{ab}+\frac13 K^2\right)
+\Lie_\omega K \right.
\nonumber\\
&&\left.\phantom{\frac11}
+4\pi\alpha\left(\rhoH+S\right) \,\right] \psi^6\sqrt{\tgamma}d^3x ,
\eeqn
where $(\Gabd - 8\pi \Tabd)\gabu=0$ is used.

In \cite{SUF04}, we have derived asymptotic conditions for 
an equality of 
the ADM mass, and the asymptotic Komar mass \cite{Komar5962}, 
$\Madm=\MK$. The equality is related to the relativistic virial 
relation for the equilibrium \cite{REMARK},   
\beq
\int x^a \gamma_a\!{}^\alpha \nabla_\beta\Tba \sqrt{-g}d^3x =0.  
\label{eq:virial}
\eeq
In the WL/NHS formulation 
the asymptotic fall-off of each field is sufficiently fast 
to enforce the equality.  And in this case, the above 
two definitions for $\MK$ agree as well.  

Finally, the total angular momentum is calculated from a volume 
form of surface integral at spatial infinity 
\beqn
J
&:=& -\frac1{8\pi}\int_\infty \pi^a{}_b\phi^b \,dS_a
\,=\,\frac1{8\pi}\int_\infty K^a{}_b\phi^b \,dS_a.  \quad
\eeqn
To calculate $J$, we set  
the surface near the boundary of the computational domain of 
the central coordinates and use the Gauss-Stokes lemma to write 
\beqn
J
&=&  \frac1{8\pi}\int_\Sigma D_a (K^a{}_b \phi^b)\,dS
\nonumber\\
&=& \frac1{8\pi}\int_\Sigma 
\left(8\pi j_a \phi^a + K^a{}_b D_a\phi^b\right)\,dS.
\nonumber\\
&=& \frac1{8\pi}\int_\Sigma 
\left(8\pi j_a \phi^a + A^a{}_b \tD_a\phi^b + \frac2{\psi}K\phi^a\zD_a\psi\right)\,
\nonumber\\
&&\ \times \psi^6\sqrt{\tgamma}d^3x.
\eeqn
The values of $J$ listed in the tables in next section, 
are calculated from the latter formula.

\section{Selected solution sequences}
\label{appsec:seq}

Selected waveless solutions of irrotational BNS 
for parametrized EOS presented in Table \ref{tab:EOS} of 
Sec.~\ref{sec:QEsol} are tabulated.  
All quantities are dimensionless 
in the geometric units $G=c=1$, except for the 
ADM mass which is in a unit of solar mass $\Madm$ [$M_\odot$]. 

\end{document}

%% file: kigou.tex
\newcommand{\Madm}{M_{\rm ADM}}
\newcommand{\MK}{M_{\rm K}}
\newcommand{\beq}{\begin{equation}} 
\newcommand{\eeq}{\end{equation}} 
\newcommand{\beqn}{\begin{eqnarray}} 
\newcommand{\eeqn}{\end{eqnarray}} 
\newcommand{\pa}{\partial}
\newcommand{\na}{\nabla}
\newcommand{\gab}{g^\alpha\!_\beta}
\newcommand{\gabu}{g^{\alpha\beta}}
\newcommand{\gabd}{g_{\alpha\beta}}

\newcommand{\gmabu}{\gamma^{ab}}
\newcommand{\gmabd}{\gamma_{ab}}
\newcommand{\tgmabu}{\tilde\gamma^{ab}}
\newcommand{\tgmabd}{\tilde\gamma_{ab}}
\newcommand{\tgamma}{\tilde\gamma}

\newcommand{\hijd}{h_{ij}}

\newcommand{\tbeta}{\tilde{\beta}}

\newcommand{\Aabd}{A_{ab}}
\newcommand{\Aba}{A_a{}\!^b}

\newcommand{\tAabd}{\tilde{A}_{ab}}

\newcommand{\tAba}{\tilde A_a{}\!^b}

\newcommand{\albe}{{\alpha\beta}}

\newcommand{\Tabd}{T_{\alpha\beta}}
\newcommand{\Tabu}{T^{\alpha\beta}}
\newcommand{\Tab}{T^\alpha{}\!_\beta}
\newcommand{\Tba}{T_\alpha{}^\beta}

\newcommand{\Gabd}{G_{\alpha\beta}}

\newcommand{\Rab}{R^\alpha{}\!_\beta}

\newcommand{\tR}{{}^{3}\!R}
\newcommand{\ttR}{{}^{3}\!\tilde R}
\newcommand{\Rnl}{\tilde{R}^{\rm NL}}
\newcommand{\Rd}{\tilde{R}^{\rm D}}

\newcommand{\tD}{\tilde D}
\newcommand{\zD}{{\raise1.0ex\hbox{${}^{\ \circ}$}}\!\!\!\!\!D}
\newcommand{\alone}{{\raise0.5ex\hbox{${}^{\ 1}$}}\!\!\!\!\alpha}
\newcommand{\Od}{{O}}

\newcommand{\tS}{\tilde S}

\newcommand{\dl}{\delta}

\newcommand{\Dl}{\Delta}

\newcommand{\Lie}{\mbox{\pounds}}

\newcommand{\compa}{M_1/R}

\newcommand{\nalam}{\mathrel{\raise0.9ex\hbox{$^\lambda$}\mkern-14mu
\lower0.0ex\hbox{$\nabla$}}}

\newcommand{\dis}{\displaystyle}
\newcommand{\gmaa}{\gamma_a\!{}^\alpha}
\newcommand{\gmbb}{\gamma_b{}^\beta}

\newcommand{\gmcdu}{\gamma^{cd}}
\newcommand{\gmcdd}{\gamma_{cd}}
\newcommand{\tgmcdu}{\tilde\gamma^{cd}}
\newcommand{\tgmcdd}{\tilde\gamma_{cd}}
\newcommand{\tomega}{\tilde\omega}
\newcommand{\tphi}{\tilde\phi}
\newcommand{\Sabd}{S_{ab}}
\newcommand{\rhoH}{\rho_{\rm H}}
\newcommand{\tC}{{\tilde C}}

\newcommand{\tbR}{{}^{3}\!\bar R}

\newcommand{\tA}{\tilde A}

\newcommand{\bD}{\bar D}
\newcommand{\Kabd}{K_{ab}}

\newcommand{\zeroD}{{\raise1.0ex\hbox{${}^{\ \circ}$}}\!\!\!\!\!D}
\newcommand{\zLap}{{\raise1.0ex\hbox{${}^{\ \circ}$}}\!\!\!\!\Delta}
\newcommand{\zna}{{\raise1.0ex\hbox{${}^{\ \circ}$}}\!\!\!\!\!\nabla}
\newcommand{\zS}{{\raise1.0ex\hbox{${}^{\ \circ}$}}\!\!\!\!\!S}
\newcommand{\tLap}{{\tilde \Delta}}

\newcommand{\SHam}{{\cal S}_{\rm H}}
\newcommand{\SMom}{{\cal S}}
\newcommand{\Str}{{\cal S}_{\rm tr}}

